\documentclass[journal]{IEEEtran}  
\usepackage{graphicx}
\usepackage[utf8]{inputenc}
\usepackage[T1]{fontenc}
\usepackage{mathtools}
\usepackage{inputenc}
\usepackage[english]{babel}
\usepackage{amsmath}
\usepackage{amssymb}
\usepackage{amsthm}
\usepackage{comment}
\usepackage{amssymb}
\usepackage{bm}
\usepackage{psfrag}
\usepackage{esdiff}
\usepackage{float}
\usepackage{caption}
\usepackage{subcaption}
\usepackage{bm}
\usepackage{algorithmicx}
\usepackage{algpseudocode}
\usepackage{lipsum}
\usepackage[linesnumbered,ruled,vlined]{algorithm2e}
\usepackage{array}
\usepackage{blindtext}
\usepackage{epstopdf}
\usepackage{ragged2e}
\usepackage{mathrsfs}
\usepackage{tabu}
\graphicspath{ {figures/} }

\usepackage{url}
\usepackage{nomencl}
\usepackage{cite}
\usepackage{lmodern}
\usepackage{enumerate}
\usepackage{tikz}
\usepackage[switch]{lineno}
\usepackage{psfrag}
\usepackage{pstool}
\usepackage{slashbox}
\usepackage{multirow}
\usepackage{hyperref}

\IEEEoverridecommandlockouts                              

\DeclareMathOperator*{\argmin}{arg\,min}
\DeclareMathOperator*{\argmax}{arg\,max}

\DeclarePairedDelimiter{\floor}{\lfloor}{\rfloor}

\newcommand{\be}{\begin{equation}}
	\newcommand{\ee}{\end{equation}}
\newcommand{\bse}{\begin{subequations}}
	\newcommand{\ese}{\end{subequations}}
\newcommand{\bewn}{\begin{equation*}}
	\newcommand{\eewn}{\end{equation*}}
\newcommand{\bbmat}{\begin{bmatrix}} 
	\newcommand{\ebmat}{\end{bmatrix}}
\newcommand{\bd}{\begin{displaymath}}
	\newcommand{\ed}{\end{displaymath}}
\newcommand{\bea}{\begin{eqnarray}}
	\newcommand{\eea}{\end{eqnarray}}
\newcommand{\ba}{\begin{array}}
	\newcommand{\ea}{\end{array}}
\newcommand{\baa}{\begin{array}{ll}}
	\newcommand{\eaa}{\end{array}}
\newcommand{\ds}{\displaystyle}
\newcommand{\bc}{\begin{center}}
	\newcommand{\ec}{\end{center}}
\newcommand{\ben}{\begin{enumerate}}
	\newcommand{\een}{\end{enumerate}}
\newcommand{\bi}{\begin{itemize}}
	\newcommand{\ei}{\end{itemize}}
\newcommand{\bt}{\begin{tabular}}
	\newcommand{\et}{\end{tabular}}
\newcommand{\bte}{\begin{table}}
	\newcommand{\ete}{\end{table}}
\newcommand{\bal}{\begin{align}}
	\newcommand{\eal}{\end{align}}

\renewcommand{\baselinestretch}{0.995}

\newcommand{\norm}[1]{\left\lVert#1\right\rVert}   

\makeatletter
\renewcommand\paragraph{\@startsection{paragraph}{4}{\z@}%
	{-2.5ex\@plus -1ex \@minus -.25ex}%
	{1.25ex \@plus .25ex}%
	{\normalfont\normalsize\bfseries}}
\makeatother



\newtheorem{remark}{Remark}

\newtheorem{theorem}{Theorem}

\newtheorem{lemma}[theorem]{\textbf{Lemma}}
\newtheorem{assumption}{\textbf{Assumption}}
\newtheorem{problem}{\textbf{Problem}}
\newtheorem{definition}{\textbf{Definition}}

\newcommand{\bR}{\mathbb{R}}
\newcommand{\bZ}{\mathbb{Z}}

\newcommand{\calA}{\mathcal{A}}
\newcommand{\calB}{\mathcal{B}}
\newcommand{\calC}{\mathcal{C}}
\newcommand{\calD}{\mathcal{D}}
\newcommand{\calE}{\mathcal{E}}
\newcommand{\calF}{\mathcal{F}}
\newcommand{\calG}{\mathcal{G}}

\newcommand{\calO}{\mathcal{O}}
\newcommand{\calP}{\mathcal{P}}

\newcommand{\calR}{\mathcal{R}}
\newcommand{\calS}{\mathcal{S}}

\newcommand{\calT}{\mathcal{T}}

\newcommand{\calV}{\mathcal{V}}
\newcommand{\calW}{\mathcal{W}}
\newcommand{\calX}{\mathcal{X}}

\newcommand{\calZ}{\mathcal{Z}}

\newcommand{\scriptA}{\mathscr{A}}
\newcommand{\scriptD}{\mathscr{D}}

\newcommand{\scriptF}{\mathscr{F}}
\newcommand{\scriptR}{\mathscr{R}}

\newcommand{\abs}[1]{\left |#1\right |}

\begin{document}
	\title{\LARGE \bf
		Aerial Swarm Defense using Interception and Herding Strategies
	}

	\author{Vishnu S. Chipade and Dimitra Panagou
		\thanks{The first author is with the Department of Aerospace Engineering,
			University of Michigan, Ann Arbor, MI, USA; the second author is with the Department of Robotics and the Department of Aerospace Engineering, University of Michigan, Ann Arbor, MI, USA.
			{\tt\small (vishnuc,
				dpanagou)@umich.edu}}
		\thanks{This work has been funded by the Center for Unmanned Aircraft Systems (C-UAS), a National Science Foundation Industry/University Cooperative Research Center (I/UCRC) under NSF Award No. 1738714 along with significant contributions from C-UAS industry members.}
	}	
	
	\maketitle
	\thispagestyle{empty}
	\pagestyle{empty}

	\begin{abstract}
		
		This paper presents a {multi-mode} solution to the problem of defending a circular protected area (target) from a wide range of attacks by swarms of \textit{risk-taking} and/or \textit{risk-averse} attacking agents (attackers). The proposed {multi-mode solution combines two defense strategies, namely: 1) an interception strategy for a team of defenders to intercept multiple \textit{risk-taking} attackers while ensuring that the defenders do not collide with each other, 2) a herding strategy to herd a swarm of \textit{risk-averse} attackers to a safe area.} 
		In particular, we develop mixed integer programs (MIPs) and geometry-inspired heuristics to distribute and assign and/or reassign the defenders to interception and herding tasks under different spatiotemporal behaviors by the attackers such as splitting into smaller swarms to evade defenders easily or high-speed maneuvers by some risk-taking attackers to maximize damage to the protected area. We provide theoretical as well as numerical comparison of the computational costs of these MIPs and the heuristics, and demonstrate the overall approach in simulations.
		
		
	\end{abstract}
	
	\section{Introduction}
	\subsection{Motivation}
	
Swarm technology has a wide range of applications \cite{brambilla2013swarm}, however may also pose threat to safety-critical infrastructure such as government facilities, airports, and military bases. The presence of adversarial agents or swarms nearby such entities, with the aim of causing physical damage or collecting critical information, can lead to catastrophic consequences. The adversarial agents (attackers) could be either risk-averse (self-interested), or risk-taking. Risk-averse attackers will try to avoid collision with other static or dynamic agents in order to avoid any damage to themselves. Risk-averse attackers could be more interested in collecting critical information by loitering around the safety-critical area (protected area) than intending to physically damage the protected area. On the other hand, risk-taking attackers will have low priority for their own survival compared to their mission. Such attackers could be interested in physically damaging the protected area. The degree of risk-aversion could vary among the attackers. Furthermore, the attackers may 1) cooperate among themselves and stay together as a swarm or do not stay together, or 2) do not cooperate among themselves.

Research has attributed various defense strategies to defend against different types of attackers, for example, 1) physical interception strategies~\cite{lee2016model,chen2017multiplayer,pierson2016intercepting,coon2017control,yan2019optimal,yan2019maximum,shishika2019team,garcia2019strategies,shishika2020cooperative,garcia2020optimal,chipade2021idcais} (mostly for risk-taking attackers), 2) herding strategies \cite{paranjape2018robotic,pierson2018controlling,haque2011biologically,varava2017herding,licitra2017single,licitra2018single,deptula2018single,nardi2018game,chipade2019herdingswarm,chipade2020herdingswarm,chipade2020multiswarm} (mostly against risk-averse attackers).
With a wide range of potential behaviors by the attackers, a single type of defense approach may not be sufficient, economical or even desirable. In this paper, we combine interception-based and herding-based defense strategies for the defenders to provide a {multi-mode} defense solution against a wide range of adversarial attacks.


	\subsection{Related work}
	{\subsubsection{Multi-player pursuit evasion games} In pursuit-evasion games a team of pursuers aims to capture or intercept a team of evaders, while the evaders aim to evade from pursuers for as long as possible. Various approaches including optimal control techniques \cite{ho1965differential}, area-minimization techniques \cite{huang2011guaranteed,pierson2016intercepting}, value function based technique \cite{jang2005control}, mean-field approach and reinforcement-learning techniques \cite{wang2020cooperative, zhou2022decentralized} exist in the literature to solve pursuit-evasion games.}
	{The existing solutions provide useful insights, however they in principle do not consider an area under risk that is targeted by the attackers. Therefore, pursuit-evasion approaches are less suitable for the class of area-defense problems studied in this paper. }

	\subsubsection{Multi-agent area (target) defense} 
	The area or target defense problem with a single agent on either team has been studied as a zero-sum differential game using various solution techniques including optimal control \cite{isaacs1999differential,venkatesan2014target,pachter2017differential,harris2020abnormal,mohanan2020target} and reachability analysis \cite{huang2011differential}. However, extending these approaches to multi-agent settings suffers from the curse of dimensionality. To remedy this, researches have been using a ``divide and conquer" approach, i.e., solve the one-on-one problem or the problem with small number of agents for all such combinations of the agents, and scale up this solution to the original multi-agent problem.
		
    In \cite{chen2017multiplayer}, the authors consider a multiplayer reach-avoid game. The authors solve the reach-avoid game for each pair of defender and attacker operating in a compact domain with obstacles using a Hamilton-Jacobi-Issacs (HJI) reachability approach. The solution is then used to assign defenders against the attackers using graph-theoretic maximum matching.
	
	In the perimeter defense problem studied in \cite{shishika2020cooperative} defenders are restricted to move on the perimeter of a protected area. Local games between small teams of defenders and attackers are solved and then assignments are done using a polynomial time algorithm.
	
	
	
	The aforementioned studies provide useful insights to the area or target defense problem, however, are limited in application due to the use of simple motion models, such as single integrators. In \cite{coon2017control}, Target-Attacker-Defender (TAD) game with agents moving under double-integrator dynamics is considered. Due to the increased computational complexity of solving a zero-sum differential game optimally for high-dimensional systems, the authors use an isochrones method to design time-optimal control strategies for the players in 1-vs-1 TAD game. However, despite bounded acceleration inputs, no bounded velocities for the agents can be ensured or is assumed in \cite{coon2017control}.
	
In all of the aforementioned work, the defenders coordinate with each other for the assignment task to intercept the attackers, however, they do not consider collision avoidance among themshelves. {Furthermore, the aforementioned interception strategies, while useful against \textit{risk-taking} attackers, may be an extreme measure against \textit{risk-averse} attackers. In other words, there may be cases where one may prefer to herd the \textit{risk-averse} defenders to some safe area and take control of these attackers in favor of the defenders, instead of intercepting them.}
	
	\subsubsection{Swarm herding}
	Herding has been studied previously in \cite{paranjape2018robotic,pierson2018controlling,haque2011biologically}. The approach in \cite{paranjape2018robotic} uses an $n$-wavefront algorithm to herd a flock of birds away from an airport, where the birds on the boundary of the flock are influenced based on the locations of the airport and a safe area.
	
	The herding method in \cite{pierson2018controlling} utilizes a circular-arc formation of herders to influence the nonlinear dynamics of the herd based on a potential-field approach, and designs a point-offset controller to guide the herd close to a specified location. 
	In \cite{haque2011biologically}, biologically-inspired strategies are developed for confining a group of agents; the authors develop strategies based on the ``wall'' and ``encirclement'' methods that dolphins use to capture a school of fish. In addition, they compute regions from which this confinement is possible; however, the results are limited to constant-velocity motion. A similar approach called \textit{herding by caging} is adopted in \cite{varava2017herding}, where a cage of high potential is formed around the attackers. An RRT approach is used to find a motion plan for the agents; however, the cage is assumed to have already been formed around the agents, while the caging of the agents thereafter is only ensured with constant velocity motion under additional assumptions on the distances between the agents. Forming such a cage could be more challenging in case of self-interested, risk-averse attackers under non-constant velocity motion.
	
	In \cite{licitra2017single,licitra2018single}, the authors discuss herding using a switched-system approach; the herder (defender) chases targets (evaders/attackers) sequentially by switching among them so that certain dwell-time conditions are satisfied to guarantee stability of the resulting trajectories. However, the assumption that only one of the targets is influenced by the herder at any time might be limiting and non-practical in real applications. 
	The authors in \cite{deptula2018single} use approximate dynamic programming to obtain suboptimal control policies for the herder to chase a target agent to a goal location. A game-theoretic formulation is used in \cite{nardi2018game} to address the herding problem by constructing a virtual barrier similar to \cite{pierson2018controlling}. However, the computational complexity due to the discretization of the state and control-action spaces limits its applicability.	
	
	Most of the aforementioned approaches for herding are limiting due to one or many of the following aspects: 1) simplified motion models, 2) absence of obstacles in the environment, 3) no consideration of inter-agent collisions, 4) assumption of a particular form of potential field to model the repulsive motion of the attackers with respect to the defenders.

	We have addressed the above issues in our recent work \cite{chipade2019herdingswarm,chipade2020herdingswarm}, which develops a method, termed as `StringNet Herding', for defending a protected area from a swarm of attackers in a 2D obstacle environment. In `StringNet Herding', a closed formation of strings (`StringNet') is formed by the defenders to surround the swarm of attackers. It is assumed that the attackers will stay together within a circular footprint as a swarm and collectively avoid the defenders. It is also assumed that the string between two defenders serves as a barrier through which the attackers cannot escape (e.g., a physical straight-line barrier, or some other mechanism). The StringNet is then controlled to herd the swarm of attackers to a safe area. The control strategy for the defenders in `StringNet Herding' is a combination of time-optimal control actions and finite-time, state-feedback, bounded control actions, so that the attackers can be herded to safe area in a timely manner. 
	
	In \cite{chipade2020multiswarm,chipade2021aerial}, we extended the `StringNet Herding' approach to scenarios where attackers no longer stay together and may split into smaller swarms in reaction to the defenders' presence. Particularly, we first identify the spatial distributions (clusters/swarms) of the attackers that satisfy certain properties, using the density-based spatial clustering for applications with noises (DBSCAN) algorithm \cite{ester1996density}. Then, we developed a mixed-integer quadratically constrained program (MIQCP) to distribute and assign the sub-teams of the defenders to the identified clusters of the attackers, so that the clusters of the attackers are herded to one of the safe areas. Note that we use swarm and cluster interchangeably throughout the paper.

	\subsection{Overview of the proposed approach}

	 {As discussed above, a wide range of approaches exist for area defense scenarios. However, only a specific type of behavior by the attackers is considered in each of the aforementioned works. To address a wide range of behaviors by the attackers a multi-mode solution is provided in this paper.} {We first make the following assumption.
	 \begin{assumption}[Inter-Defender Collision-Aware Interception Strategy (IDCAIS)]
	 	There exists an interception strategy to intercept multiple attackers in an area-defense game, such that the defenders account for inter-defender collisions while they intercept the attackers as quickly as possible.
	 \end{assumption}
 Such interception strategy is provided in \cite{chipade2021idcais} (under review).}

\begin{figure}
		\centering
		\includegraphics[width=\linewidth,]{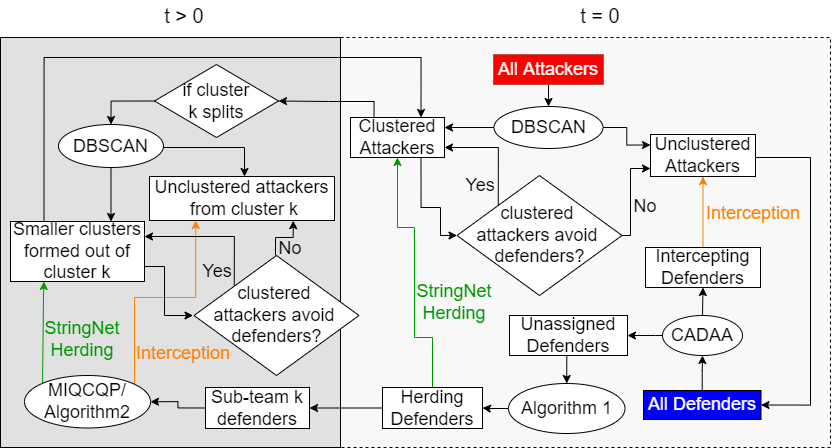}
		\caption{{Overview of the Multi-mode Defense Approach}}
		\vspace{-2mm}
		\label{fig:multimodeStrategyBlockDiagram}
	\end{figure}

 {
	 The multi-mode defense approach discussed in this paper is summarized in Figure \ref{fig:multimodeStrategyBlockDiagram}.
	 In this multi-mode defense approach,} the spatial distributions of the attackers are continuously monitored using the DBSCAN algorithm, which classifies attackers into clusters of at least three agents. The attackers that either belong to clusters of less than 3 attackers, or are classified as noises by the DBSCAN algorithm, are called unclustered attackers. At time $t = 0$ s {(the right half section in Figure \ref{fig:multimodeStrategyBlockDiagram})}, the defending team employs the IDCAIS against the unclustered attackers; under this interception strategy, some of the defenders are assigned to intercept the unclustered attackers in minimum time using collision-aware defender-to-attacker assignment (CADAA) \cite{chipade2021idcais} {(discussed later)}, these defenders are called intercepting defenders. The rest unassigned defenders, called herding defenders, are distributed into sub-teams and assigned to herd the identified clusters of the attackers to one of the safe areas using `StringNet Herding' approach \cite{chipade2020multiswarm}, as long as the attackers stay together and avoid the defenders. If the attackers further split into new smaller clusters and/or individual attackers (unclustered attackers) {at some time $t>0$ (shown in the left half section in Figure \ref{fig:multimodeStrategyBlockDiagram}), then the defenders are also further distributed into smaller sub-teams and assigned to herd the newly formed attackers' clusters and to intercept the newly-identified unclustered attackers that separated from the original cluster of the attackers using an optimal assignment algorithm}. 

	\subsection{Summary of our contributions}
	We develop a {multi-mode} defense strategy against wide range of swarm attacks using the IDCAIS and the `StringNet Herding' \cite{chipade2020herdingswarm} approach. Compared to the prior literature and our own work, the contributions of this paper are:
	\bi
	\item[1)] a centralized, iterative algorithm to assign the defenders to the attackers' clusters identified at $t=0$ so that the defenders gather on the shortest paths of the attackers' swarms to the protected area before the attackers reach there;
	\item[2)] a decentralized algorithm using mixed integer quadratically constrained quadratic programs (MIQCQPs) to assign the defenders to intercept the unclustered attackers, and to herd the attackers' newly-formed swarms in the case a swarm of attackers splits into smaller swarms  at any future time $t>0$;
	\item[3)] heuristics to solve the MIQCQP approximately but quickly to find the assignment in real time;
	\item[4)] theoretical as well as numerical comparison of the computational cost of the assignment algorithms.
	\ei
	
	\subsection{Organization}
	The rest of the paper is structured as follows. {Section \ref{sec:math_model} provides the mathematical modeling, assumptions made and a statement of the problem studied.}
	The strategy and the assignment algorithms of the {multi-mode} defense approach are discussed in Section \ref{sec:multimode_defense_strategy}.{More specifically, the optimal assignment algorithms at $t=0$ and $t>0$, their sub-optimal but computationally better alternative algorithms, heuristics to solve these optimal assignment problems in a computationally efficient manner, as well as their performance comparison are discussed in Section~\ref{sec:multimode_defense_strategy}. Simulation results for various scenarios demonstrating the proposed {multi-mode} framework are provided in Section \ref{sec:simulations}.} The paper is concluded in Section \ref{sec:conclusions}.

	\section{Modeling and Problem Statement}\label{sec:math_model}
	\textit{Notation}: We use $\norm{\cdot}$ to denote the Euclidean norm of its argument. $\abs{\cdot}$ denotes absolute value of a scalar argument or cardinality of a set argument. A ball of radius $\rho$ centered at the origin is defined as $\calB_{\rho}=\{\mathbf{r} \in \bR^2| \norm{\mathbf{r}}\le\rho \}$ and that centered at $\mathbf{r}_c$ is defined $\calB_{\rho}(\mathbf{r}_c)=\{\mathbf{r} \in \bR^2| \norm{\mathbf{r}-\mathbf{r}_c}\le\rho \}$.
	 $A\backslash B$ denotes all the elements of the set $A$ that are not in the set $B$. Some most commonly used variables in the paper are described in Table~\ref{tab:notation}.

	\begin{table}[h!]
    \begin{center}
    \caption{Table of notation}
    \label{tab:notation}
    \begin{tabular}{l@{\hskip 1mm} l} 
    $\calA_i$ & denotes the $i^{th}$ attacker\\
    $\calA_{c_k}(t)$ & denotes the group of attackers indexed by $A_{c_k}(t)$\\
    $A_{c_k}(t)$ & set of indices of the attackers in $k^{th}$ cluster \\
    & of attackers at time $t$\\
    $A_{uc}(t)$ & set of indices of the unclustered attackers at time $t$\\
    $A_{c}^{(k)}(t)$ & set of indices of clusters of the attackers separated \\
    & from the $k^{th}$ cluster of attackers at time $t$\\
    $A_{uc}^{(k)}(t)$ & set of indices of the unclustered attackers separated \\
    & from the $k^{th}$ cluster of attackers at time $t$\\
    $\scriptA_k(t_{se})$ & data structure storing information of the attackers\\
    & in $k^{th}$ cluster of attackers after it splits at $t=t_{se}$\\
    $\scriptA_k(t_{se}).f$ & denotes the data field $f$ of the data structure\\
    &  $\scriptA_k(t_{se})$ at time $t=t_{se}$ \\
    $\calD_j$ & denotes the $j^{th}$ defender\\
    $\calD_{c_k}(t)$ & denotes the group of defenders indexed by $D_{c_k}(t)$\\
    $\calD_{c_k}^c(t), \; \calD_{c_k}^t(t)$ & group of central and terminal defenders on the\\
    & Open-StringNet $\calG_{sn}^{op}(D_{c_k}(t))$, resp. \\
    $\calD_{c_k}^l(t), \; \calD_{c_k}^r(t)$ & group of terminal defenders on the left and  right \\
    &  end of Open-StringNet $\calG_{sn}^{op}(D_{c_k}(t))$, resp. \\
    $D_{c_k}(t)$ & set of indices of the defenders assigned to $k^{th}$ \\
    & cluster of attackers at time $t$\\
    $\scriptD_k(t_{se})$ & data structure storing information of the defenders\\
    & indexed by $D_k(t_{se}^-)$\\
    $\scriptD_k(t_{se}).f$ & denotes the data field $f$ of the data structure\\
    &  $\scriptD_k(t_{se})$ at time $t=t_{se}$ \\
    $\calG_{sn}^{cl}(I_d)$ & Closed-StringNet formed by the defenders with\\
    & indices as in the set $I_d$\\
    $\calG_{sn}^{op}(I_d)$ & Open-StringNet formed by the defenders with \\
    & indices as in the set $I_d$\\
    $I_a,\; I_{ac}(t)$ & equals set $\{1,2,...,N_a\}$, $\{1,2,...,N_{ac}(t)|)\}$, resp.\\
    $I_d, \; I_{dc_k}(t)$ & equals set $\{1,2,...,N_d\}$, $\{1,2,...,\scriptR_d(|A_{c_k}(t)|)\}$, resp. \\
    $N_a,\; N_{ac}(t)$ & number of attackers and attackers' clusters, resp.\\
    $N_d,$ & number of defenders\\
    $\scriptR_d(\cdot)$ & defender-to-attacker resource allocation function\\
    $\mathbf{r}_{ai}, \;\mathbf{r}_{dj}$ & position of $i^{th}$ attacker, $j^{th}$ defender, resp. \\
    $\mathbf{r}_{sm}$ & center $m^{th}$ safe area\\
    $t_{se}$ & time at which attackers' split event happens\\
    $t_{se}^-$ & time instant just before attackers' split event\\
       $\mathbf{u}_{ai}, \;\mathbf{u}_{dj}$ & acceleration of $i^{th}$ attacker, $j^{th}$ defender, resp. \\
       $\mathbf{v}_{ai}, \;\mathbf{v}_{dj}$ & velocity of $i^{th}$ attacker, $j^{th}$ defender, resp. \\
       $\beta_{c}(t)$ & set of mappings of defenders' assignment to the \\
       & clusters of the attackers at time $t$\\
       $\beta_{c_k}(t)$ & mapping that assigns defenders to the $k^{th}$ cluster\\
       &  of the attackers at time $t$\\
       $\beta_{uc}(t)$ & mapping that assigns defenders to the unclustered\\
       & attackers at time $t$\\
       $\rho_{d}^{int}$ & interception radius of a defender\\
       $\delta_{jk}^{herd}(t)$ & decision variable to decide if $\calD_{j}$ is assigned to\\
       & herd attackers' swarm $\calA_{c_k}$ at time $t$\\
       $\delta_{ji}^{int}(t)$ & decision variable to decide if $\calD_{j}$ is assigned to\\
        & intercept the attacker $\calA_{i}$ at time $t$\\
      \end{tabular}
    \end{center}
    \end{table}

\begin{figure}
		\centering
		\includegraphics[width=.88\linewidth,trim={10cm 2.8cm 8.5cm 2.1cm},clip]{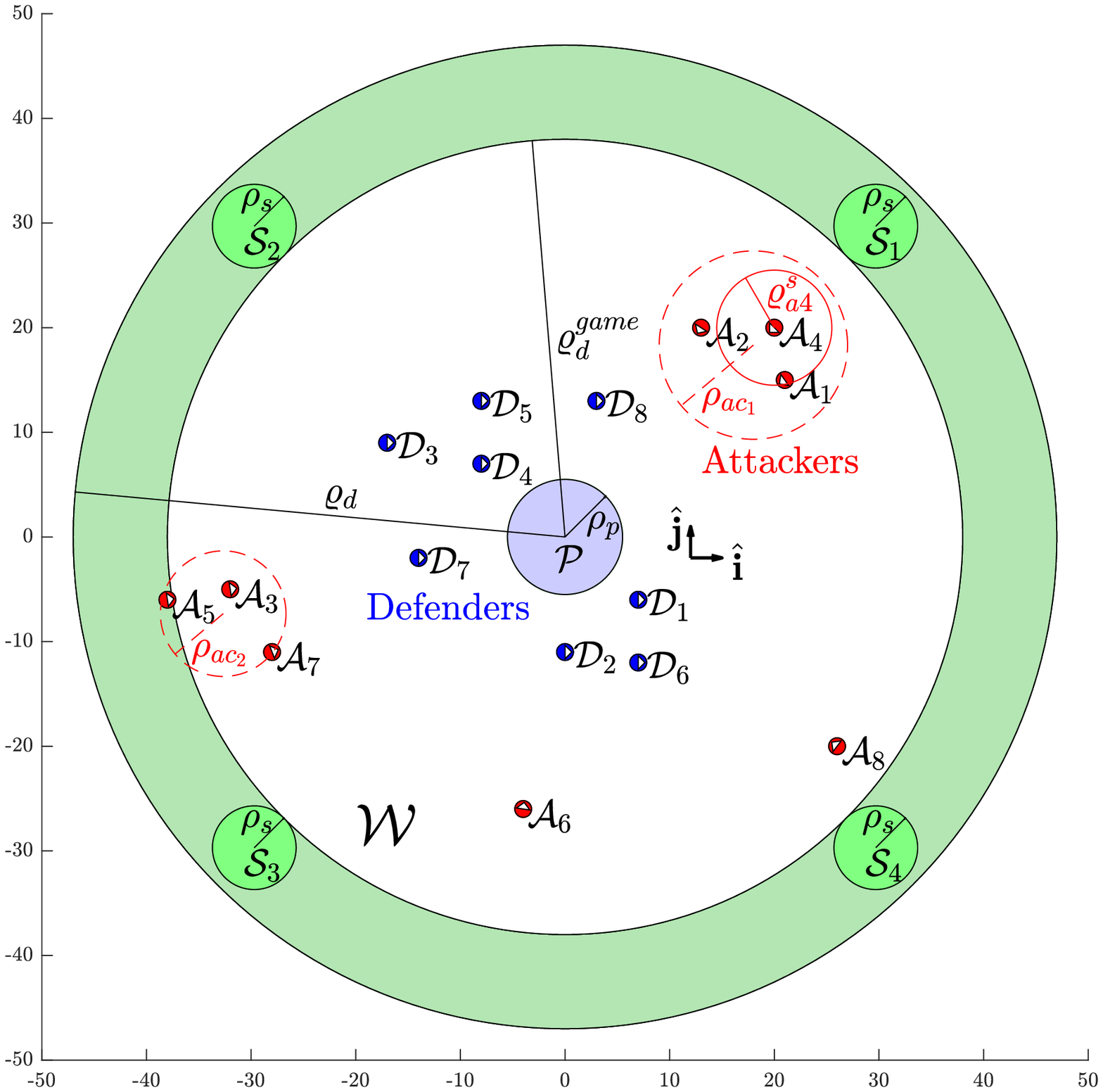}
		\vspace{-1mm}
		\caption{{Schematic of a scenario showing multiple attackers (red filled circles with white arrows), some as risk-averse swarms while some individual risk-taking attackers, trying to reach the protected area $\calP$ and defenders (blue filled circles with white arrows) spread around $\calP$.}}
		\vspace{-2mm}
		\label{fig:problem_formulation}
	\end{figure}

We consider $N_a$ attackers denoted as $\calA_i$, $i \in I_a= \{1,2,...,N_a\}$, and $N_d$ defenders denoted as $\calD_j$, $j \in I_d= \{1,2,...,N_d\}$, operating in a 2D environment $\calW \subseteq \mathbb{R}^2$ that contains a protected area $\calP \subset \calW$, defined as $\calP=\{\mathbf{r} \in \bR^2 \;| \; \norm{\mathbf{r}}\le \rho_p\}$, and $N_s$ safe areas $\calS_{m} \subset \calW$, defined as $\calS_{m}=\{\mathbf{r} \in \bR^2 \; | \; \norm{\mathbf{r}-\mathbf{r}_{sm}}\le \rho_{sm}\}$, for all $m \in I_s=\{1,2,...,N_s\}$, where $\rho_p$ and $\rho_{sm}$ are the radii of the protected area and $m^{th}$ safe area, respectively, and $\mathbf{r}_{sm}$ is the center of $m^{th}$ safe area. {Visual depiction of the above elements is shown in Figure~\ref{fig:problem_formulation}}. 
	The number of defenders is no less than that of attackers, i.e., $N_d \ge N_a$. The agents $\calA_i$ and $\calD_j$ are modeled as discs of radii $\rho_a$ and $\rho_d$, where $ \rho_d \le \rho_a$, respectively.	Let $\mathbf{r}_{ai}=[x_{ai}\; y_{ai}]^T$ and $\mathbf{r}_{dj}=[x_{dj}\; y_{dj}]^T$ be the position vectors of $\calA_i$ and $\calD_j$, respectively; $\mathbf{v}_{ai}=[v_{x_{ai}}\; v_{y_{ai}}]^T$, $\mathbf{v}_{dj}=[v_{x_{dj}}\; v_{y_{dj}}]^T$ be the velocity vectors, respectively, and $\mathbf{u}_{ai}=[u_{x_{ai}}\; u_{y_{ai}}]^T$, $\mathbf{u}_{dj}=[u_{x_{dj}}\; u_{y_{dj}}]^T$ be the accelerations, which serve also as the control inputs, respectively, all resolved in a global inertial frame $\calF_{gi} (\hat {\mathbf{i}}, \hat {\mathbf{j}})$ (see Fig.\ref{fig:problem_formulation}).
	The agents move under double integrator (DI) dynamics with linear drag (damped double integrator), similar to isotropic rocket \cite{bakolas2014optimal}: 
	
	\be \label{eq:dampedDIDyn}
	\baa
	\dot{\mathbf{x}}_{\star}=\bbmat \dot{\mathbf{r}}_{\star}\\\dot{\mathbf{v}}_{\star} \ebmat  = \bbmat \mathbf{0}_{2} & \mathbf{I}_2 \\ \mathbf{0}_{2} & -C_D\mathbf{I}_2 \ebmat \mathbf{x}_{\star} + \bbmat \mathbf{0}_2\\ \mathbf{I}_2\ebmat \mathbf{u}_{\star}
	\eaa
	\ee
	where ${\star} \in \{ai| i \in I_a\} \cup \{dj | j \in I_d \}$, $C_D>0$ is the known, constant drag coefficient. The accelerations $\mathbf{u}_{ai}$ and $\mathbf{u}_{dj}$ are bounded by $\bar{u}_a $, $\bar{u}_d$ as given in \eqref{eq:constraints} such that $\bar{u}_a < \bar{u}_d$.
	\be\label{eq:constraints}
	\baa
	\norm{\mathbf{u}_{ai}} \le \bar{u}_a, \quad
	\norm{\mathbf{u}_{dj}} \le \bar{u}_d, 
	\eaa
	\ee
	By incorporating the drag term, the damped double integrator \eqref{eq:dampedDIDyn} inherently poses a speed bound on each agent under a limited acceleration control, i.e., $\norm{\mathbf{v}_{ai}}<\bar{v}_a=\frac{\bar{u}_a}{C_D}$ and $\norm{\mathbf{v}_{dj}}<\bar{v}_d=\frac{\bar{u}_d}{C_D}$, and does not require an explicit constraint on the velocity of the agents while designing bounded controllers, as in earlier literature.
	So we have $\mathbf{x}_{ai} \in \calX_a$, for all $i \in I_a$, where $\calX_a = \bR^2 \times \calB_{\bar{v}_a}$ and  $\mathbf{x}_{dj} \in \calX_d$, for all $j \in I_d$, where $\calX_d = \bR^2 \times \calB_{\bar{v}_d}$. We make the following assumption:
	\begin{assumption} \label{assum:measurements}
	All the defenders know the position $\mathbf{r}_{ai}$ and velocity $\mathbf{v}_{ai}$ of the attacker $\calA_i$ that lies inside a circular sensing zone $\calZ_d=\{\mathbf r \in \mathbb{R}^2 |\; \norm{\mathbf{r}} \le \varrho_d\}$ for all $i \in I_a$, where $\varrho_d>0$ is the radius of the defenders' sensing zone. Every attacker $\calA_i$ has a similar local sensing zone $\calZ_{ai}=\{\mathbf{r} \in \bR^2 \;| \; \norm{\mathbf{r}-\mathbf{r}_{ai}}\le \varrho_{ai} \}$, where $\varrho_{ai}>0$ is the radius of $\calA_i$'s sensing zone (Fig.~\ref{fig:problem_formulation}).
	\end{assumption}
	For Assumption \ref{assum:measurements} to hold, a system of sensors such as radars, lidars, cameras, etc., that are spatially distributed around the protected area can be used. {The data from these sensors are assumed to be processed by a central computer and communicated to all the defenders.}
	
Each defender is capable of connecting to other two defenders via string barriers. String barriers are realized as impenetrable and extendable line barriers (e.g., spring-loaded pulley and a rope or other similar mechanism \cite{mirjan2016building}) that prevent attackers from passing through them. The extendable string barrier allows free relative motion of the two defenders connected by the string. The string barrier can have a maximum length of $\bar{R}_{sb}$. If the string barrier were to be physical one, then it can be established between two defenders $\calD_j$ and $\calD_{j'}$ only when they are close to each other and have almost same velocity, i.e., $\norm{\mathbf{r}_{dj}-\mathbf{r}_{dj'}}\le \epsilon_1<\bar{R}_{sb}$ and $\norm{\mathbf{v}_{dj}-\mathbf{v}_{dj'}}\le \epsilon_2$, where $\epsilon_1$ and $\epsilon_2$ are small numbers {that depend on the physical size of the defenders as well as the mechanism and their capability to physically connect at a given distance.}. Each defender $\calD_j$ is endowed with an interception/capture radius $\rho_d^{int}$, i.e., the defender $\calD_j$ {is able to physically damage an attacker} $\calA_i$ when $\norm{\mathbf{r}_{dj}(t)-\mathbf{r}_{ai}(t)}<\rho_d^{int}$ for some $t>0$. 

	The goal of the attackers is to send as many attackers as possible to the protected area $\calP$. The defenders aim to either intercept these attackers or herd them away to one of the safe areas in $\calS=\{\calS_1,\calS_2,...,\calS_{N_s}\}$  in order to defend the protected area $\calP$. Formally, we consider the following problem.
	
	
	\begin{problem}[Swarm Defense] \label{prob:swarm_defense}
	Design a defense strategy for a team of defenders to defend a protected area from a wide range of adversarial attacks by attackers, where attackers could possibly stay together as swarms or stay alone during the attack.
	\end{problem}
	Next, we discuss the {multi-mode} defense strategy that addresses Problem \ref{prob:swarm_defense}.

\section{{Multi-mode} Defense 
Strategy}\label{sec:multimode_defense_strategy}
The attackers may show wide range of behaviors, such as: some or all attackers staying close together, some or all attackers avoiding defenders while attacking the protected area, some attackers not intending to damage the protected area but only interested in reaching in its neighborhood maybe for collecting some key information, while some attackers only interested in physically damaging the protected area at any cost, etc. 
	
In this section, we provide a {multi-mode} algorithm to combine the `StringNet Herding' approach developed in \cite{chipade2019herdingswarm,chipade2020herdingswarm,chipade2020multiswarm} and {the IDCAIS} to defend against wide range of behaviors by the attackers discussed earlier. In the following, we first revisit some key definitions related to `StringNet Herding'.

\begin{definition}[Closed-StringNet] \label{def:closed_StringNet} The Closed-StringNet $\calG_{sn}^{cl}(I_d)= (\calV_{sn}^{cl}(I_d),$ $\calE_{sn}^{cl}(I_d))$ is a cycle graph consisting of: 1) a subset of defenders as the vertices, $\calV_{sn}^{cl}(I_d)=\{\calD_j \;|\; j \in I_d\}$, 2) a set of edges, $\calE_{sn}^{cl}(I_d)=\{(\calD_j,\calD_{j'}) \in \calV_{sn}^{cl}(I_d) \times \calV_{sn}^{cl}(I_d) | \calD_j \overset{s} \longleftrightarrow \calD_{j'} \}$, where the operator $\overset{s} \longleftrightarrow$ denotes an impenetrable line barrier between the defenders.
\end{definition}

\begin{definition}[Open-StringNet] The Open-StringNet $\calG_{sn}^{op}(I_d)= (\calV_{sn}^{op}(I_d),$ $\calE_{sn}^{op}(I_d))$ is a path graph consisting of: 1) a set of vertices, $\calV_{sn}^{op}(I_d)$ and 2) a set of edges, $\calE_{sn}^{op}(I_d)$, similar to that in Definition \ref{def:closed_StringNet}.
\end{definition}

The `StringNet Herding' approach consists of four phases: 1) gathering, 2) seeking, 3) enclosing, and 4) herding. In the gathering phase, the defenders establish an Open-StringNet on the time-optimal path of the attackers' swarm. Then in the seeking phase they seek to get close to the attackers' swarm if they had not traveled along their time-optimal trajectory as expected by the defenders. During the seeking phase, the defenders ensure that they maintain the Open-StringNet formation. Next, during the enclosing phase, as the defenders come sufficiently close to the attackers, they enclose the attackers by establishing a Closed-StringNet around the attackers' swarm. In the herding phase, the Closed-StringNet is moved to the nearest safe area which also takes the enclosed attackers to the safe area. 


Next, we describe how the defenders are assigned to either intercept the unclustered (more likely risk-taking) attackers or herd the clustered (more likely risk-averse) attackers during different temporal and spatial events.

\subsection{Optimal assignment at $t=0$}
We first identify spatial distributions (clusters) of the attackers that are detected in the annular region between the circles ${\mathbf{r}}=\varrho_d$ and ${\mathbf{r}}=\varrho_d^{game}$ (see Fig.~\ref{fig:problem_formulation}). For the cluster identification, we use DBSCAN algorithm \cite{ester1996density} with parameters $\varepsilon_{nb}=\frac{\bar{\rho}_{ac} (m_{pts}-1) }{N_a-1}$ and $m_{pts}=3$ where $\bar{\rho}_{ac} =\frac{\bar{R}_{sb}}{2} \cot(\frac{\pi}{N_d})$ is the radius of the largest circle inscribed in the largest Closed-StringNet formation that can be formed by the $N_d$ defenders. This choice of parameters for the DBSCAN algorithm ensures that the identified clusters have more than 3 attackers in them and have sizes for which subteams of the defenders can be found which can herd these clusters. This is because one needs at least 3 defenders to form a Closed-StringNet and if $N_d = N_a$ then we may not have enough defenders to enclose all swarms of the attackers with less than 3 attackers in them. Hence all the swarms of the attackers with less than 3 attackers will be termed as singular swarms and the member attackers of these singular swarms will be identified as noise by DBSCAN algorithm and classified as unclustered attackers. For more details on how the parameters of the DBSCAN are chosen, refer to \cite{chipade2020multiswarm}.
 Let $\calA_c(0)=\{\calA_{c_1}(0),\calA_{c_2}(0),\dots, \calA_{c_{N_{ac}(0)}}(0)\}$ be the set of $N_{ac}(0)$ swarms of the attackers at $t=0$ identified using the DBSCAN algorithm. Here $\calA_{c_k}(0)=\{\calA_i| i \in A_{c_k}(0)\}$, for $k \in I_{ac}(0)=\{1,2,3,...,N_{ac}(0)\}$ where $A_{c_k}(0) \subseteq I_a$ is the set of indices of the attackers that belong to the $k^{th}$ cluster of the attackers at $t=0$. Let $\calA_{uc}(0)=\{\calA_i|i \in A_{uc}(0)\}$ denote the set of unclustered attackers where $A_{uc}(0) \subseteq I_a$ is the set of indices of the attackers that are not clustered by the DBSCAN algorithm, i.e., the attackers that are treated as the noises by the DBSCAN algorithm. 
 The defenders aim to intercept the unclustered attackers assuming that these attackers are risk-taking while they attempt to herd the clustered attackers with the hope that the clustered attackers will stay together and try to avoid the defenders. For this we need to assign some individual defenders to intercept the unclustered attackers and some sub-teams of the defenders to herd the identified clusters of the attackers. Since the unclustered attackers are likely to be risk-taking and hence pose more risk to the protected area, the assignment of the best defenders to intercept these unclustered attackers is done first and then the rest defenders are assigned to herd the clustered attackers.
 
We first us collision-aware defender-to-attacker assignment (CADAA) to assign defenders to intercept the identified unclustered attackers $\calA_{uc}(0)$ such that these attackers are intercepted as quickly as possible and the possible collisions among the defenders are minimized. Let $\delta_{ji}^{int}(0)$ be the binary decision variable at time $t=0$ that takes value 1 if the defender $\calD_j$ is assigned to intercept attacker $\calA_i$ and 0 otherwise. Let $C_d^{int}(\mathbf{X}_{dj}^{ai})$ be the cost incurred by the defender $\calD_j$ to capture the attacker $\calA_i$ and is given by:
\be \label{eq:cost_to_capture_attacker}
C_d^{int}(\mathbf{X}_{dj}^{ai})= \begin{cases}
		t_d^{int}(\mathbf{x}_{dj}, \mathbf{x}_{ai}), & \text{if } \mathbf{x}_{ai} \in \calR_d(\mathbf{x}_{dj});\\
		c_l, & \text{otherwise};
\end{cases}
\ee
where  $\mathbf{X}_{dj}^{ai} = [\mathbf{x}_{dj}^T, \mathbf{x}_{ai}^T]^T$, $t_d^{int}(\mathbf{x}_{dj}, \mathbf{x}_{ai})$ is the minimum time required by the defender $\calD_j$ to capture the attacker $\calA_i$ that is moving towards the protected area $\calP$ under time-optimal control action as defined in \cite{chipade2021idcais}, $c_l \;(>>1)$ is a very large number, and $\calR_d(\mathbf{x}_{dj}) = \{\mathbf{x}_a \in  \bar{\calX}_a|t_{d}^{int}(\mathbf{x}_{d},\mathbf{x}_{a})-t_{a}^{int}(\mathbf{x}_{a},\mathbf{r}_p)\le  0\}$ is the winning region of the defender $\calD_j$ starting at $\mathbf{x}_{dj}$, where $\bar{\calX}_a=(\bR^2\backslash\calP)\times\calB_{\bar{v}_a}$, $t_{a}^{int}(\mathbf{x}_{a},\mathbf{r}_p)$ is the time that the attacker starting at $\mathbf{x}_{a}$ requires to reach the protected area at $\mathbf{r}_p$. Let
$C_d^{col}(\mathbf{X}_{dj}^{ai}, \mathbf{X}_{dj'}^{ai'})$ is the cost associated with a collision that may occur between the two defenders that are assigned interception task and is defined as:
\be
C_d^{col}(\mathbf{X}_{dj}^{ai}, \mathbf{X}_{dj'}^{ai'}) = 
\begin{cases}
\frac{1}{t_d^{col}(\mathbf{X}_{dj}^{ai}, \mathbf{X}_{dj'}^{ai'})}, & \text{if $\calD_j$ \& $\calD_{j'}$ collide};\\
0, & \text{otherwise}.
\end{cases}
\ee
where $t_d^{col}(\mathbf{X}_{dj}^{ai}, \mathbf{X}_{dj'}^{ai'})$ is time of collision between $\calD_j$ and $\calD_{j'}$ on their time-optimal trajectories.

{We find the optimal $\delta_{ji}^{int*}(0)$ by solving the following CADAA problem at $t=0$: 
\bse \label{eq:defender_attackers_assign_MIQP}
	\begin{align}
	\argmin_{\bm{\delta}^{int}(0)} & \ds \sum_{i \in A_{uc}(0)} \sum_{j \in I_d} \Bigl (({1-w})C_d^{int}(\mathbf{X}_{dj}^{ai})\delta_{ji}^{int}(0) + \nonumber\\
	&\quad {w} \ds \sum_{i' \in A_{uc}(0)} \sum_{j' \in I_d} C_d^{col}(\mathbf{X}_{dj}^{ai}, \mathbf{X}_{dj'}^{ai'})\delta_{ji}^{int}(0)\delta_{j'i'}^{int}(0)\Bigr) \label{eq:MILP_cost}\\
	\vspace{2mm}
	\text{Subject to } & \textstyle \sum_{i \in A_{uc}(0)} \delta_{ji}^{int}(0) = 1, \quad \forall j \in I_d; \label{eq:MILP_constraint_1}\\
	\vspace{2mm}
    & \textstyle \sum_{j \in I_{d}} \delta_{ji}(0) = 1, \quad \forall i \in A_{uc}(0); \label{eq:MILP_constraint_2}\\
	& \textstyle \delta_{ji}^{int}(0)\in \{0,1\}, \quad \forall j \in I_d, \; i \in A_{uc}(0);
	\end{align}
	\ese
	where $\bm{\delta}^{int}(0)=[\delta_{ji}^{int}(0)| i \in A_{uc}(0), j \in I_d]^T \in \{0,1\}^{N_d|A_{uc}(0)|}$ is the binary decision vector and  $w \in (0,1)$ is user specified weight of the collision cost that is used to adjust the importance of the collisions among the defenders and the time to intercept the attackers at the assignment stage.} 

 A mapping $\beta_{uc}(0,\cdot):\{i \in A_{uc}(0)\}\rightarrow \{j \in I_d\}$, which gives the index of the defender assigned to intercept a given unclustered attacker $\calA_i$ is then defined as:
 \be
 \beta_{uc}(t,i)= \argmax_{j} \delta_{ji}^{int*}(0), \quad \forall t\ge0.
 \ee
 Let $\calD_{uc}(0) = \{\calD_{\beta_{uc}(t,i)}| i \in A_{uc}(0)\}$ denote the set of defenders that are assigned to the unclustered attackers $\calA_{uc}(0)$ and $D_{uc}(0)=\{\beta_{uc}(t,i)| i \in A_{uc}(0)\}$ be the set of indices of the defenders in $\calD_{uc}(0)$. Let $\calD_c(0)=\{\calD_j | j \in D_c(0)\}$ denote the set of all the other unassigned defenders, where $D_{c}(0)=I_d \backslash D_{uc}(0)$. These unassigned defenders $\calD_c(0)$ are then employed to herd the identified clusters of the attackers.

Next, we describe a centralized approach to find a time-opimal, collision free motion plan for the defenders in $\calD_c(0)$ to gather on the shortest paths of the attackers' swarms. 
\subsubsection{Centralized Approach} In this approach, the two problems: i) of choosing the best gathering formations, and ii) of the assignment of the defenders in $\calD_c(0)$ to the goal locations on these gathering formations are solved simultaneously. We provide a bisection method based iterative scheme as detailed in Algorithm \ref{alg:gathering_formations_centralized} to solve the above two problems simultaneously. Let $\scriptR_d(N_a):\bZ_{>0} \rightarrow \bZ_{>0}$ be the defender-to-attacker resource allocation function that outputs the number of the defenders that can be assigned to the given $N_a$ attackers. We make the following assumption about the defender-to-attacker resource allocation function.
	\begin{assumption}\label{assum:resource_allocation}
	The defender-to-attacker resource allocation function is a strictly monotonically increasing function, i.e., $\scriptR_d(N_a)<\scriptR_d(N_a+1)$, such that $\scriptR_d(N_a)\ge N_a$.
	\end{assumption}
	Assumption~\ref{assum:resource_allocation} ensures that there are adequate number of defenders to go after each attacker in the event the attackers in the swarm disintegrate into singular swarms.

Consider a line formation $\scriptF_{dc_k}^{line}$ characterized by positions $\mathbf{p}_k^{line}(\mathbf{r}_{df_k},\phi_{k})=\{\mathbf{p}_{k,1}^{line},\mathbf{p}_{k,2}^{line},...,\mathbf{p}_{k,\scriptR_d(|A_{c_k}|)}^{line}\}$
where
\be 
\arraycolsep=1.4pt
\baa
\mathbf{p}_{k,l}^{line}(\mathbf{r}_{df_k},\phi_{k}) \triangleq \mathbf{r}_{df_k} + \hat{R}_{l} \hat{\mathbf{o}} (\phi_{k}+\frac{\pi}{2}), \quad 
\eaa
\ee
for all  $l \in I_{dc_k}(0)=\{1,2,...,\scriptR_d(|A_{c_k}(0)|)\}$, where $\hat{\mathbf{o}}(\theta)=[ \cos(\theta), \; \sin(\theta) ]^T$ is the unit vector making an angle $\theta$ with $x$-axis, and $\hat{R}_{l}= \hat{R}_{d}^{d,g} \left(\frac{\scriptR_d(|A_{c_k}|)-2l+1}{2} \right)$, where $\hat{R}_{d}^{d,g} (\le \bar{R}_{sb})$ is the user defined separation between the defenders at the gathering formation.

	 
	  Corresponding to each attackers' cluster $\calA_{c_k}$, the desired gathering formation $\scriptF_{dc_k}^g$ for the defenders to gather at is chosen to be a line formation\footnote{This is a better choice compared to a semicircular formation as chosen in \cite{chipade2020herdingswarm}. Because, the semicircular formation, for a given length constraint on the string barrier ($\bar{R}_{sb}$), creates smaller blockage to the attackers as compared to the line formation. Although, Completing a circular formation starting from a semicircular formation of the same radius is faster. It is a trade-off between effectiveness and speed.}  $\scriptF_{dc_k}^{line}$ centered at $\mathbf{r}_{df_k}$ with orientation $\phi_k$, characterized by the positions $\bm{\xi}_{c_k}^g=\{\bm{\xi}_{c_k,1}^g,\bm{\xi}_{c_k,2}^g,...,\bm{\xi}_{c_k,\scriptR_d(|A_{c_k}|)}^g\} = \mathbf{p}_k^{line}(\mathbf{r}_{df_k},\phi_{k})$, as obtained in Algorithm \ref{alg:gathering_formations_centralized}.
	 These positions are static, i.e., $\dot{\bm{\xi}}_{c_k,l}^g=\ddot{\bm{\xi}}_{c_k,l}^g=\mathbf{0}$ for all $l \in I_{dc_k}$. The gathering centers  $\mathbf{r}_{df_k}$, for all $k \in I_{ac}(0)$, are chosen to lie outside the protected area $\calP$. Algorithm \ref{alg:gathering_formations_centralized} also outputs the Defender-to-Attacker-Swarm Assignment (DASA), $\beta$, which is defined formally as:
	 \begin{definition}[Defender-to-Attacker-Swarm Assignment]
		A set $\beta_{c}(t)$ $=\{\beta_{c_1}(t,\cdot),\beta_{c_2}(t,\cdot),...\beta_{c_{N_{ac}(t)}}(t,\cdot) \}$ of mappings $\beta_{c_k}(t,\cdot): \{1,2,...,$ $\scriptR_d(|\calA_{c_k}(t)|)\} \rightarrow I_d$, where,  for all $k \in I_{ac}(t)$, $\beta_{c_k}(t,l)$ gives the index of the defender, at time $t$, that is assigned to either gather at position $\bm{\xi}_{c_k,l}^g$ on the time-optimal path of swarm $\calA_{c_k}(t)$ during the gathering phase, or track the desired position $\bm{\xi}_{c_k,l}^s$ or $\bm{\xi}_{c_k,l}^e$ or $\bm{\xi}_{c_k,l}^h$ during the seeking or enclosing or herding phase, respectively, in order to successfully herd the swarm $\calA_{c_k}(t) $ to the closest safe area.
	\end{definition}
The set of defenders assigned to gather on the path of the cluster $\calA_{c_k}(0)$ is denoted by $\calD_{c_k}(0)=\{\calD_j| j \in D_{c_k}(0)\}$, where $D_{c_k}(0)$ is the set of indices defined as: $D_{c_k}(0)=\{\beta_{c_k}(0,1), \beta_{c_k}(0,2),..., \beta_{c_k}(0,\scriptR_d(0,|\calA_{c_k}|)) \}$ for all $k \in I_{ac}(0)$. Each of these sub-teams $\calD_{c_k}(0)$'s of the defenders are tasked to achieve the Open-StringNet formations $\calG_{sn}^{op}(D_{c_k}(0))$ on the shortest paths of the oncoming attacking swarms. Assuming $N_d=N_a$, we choose $\scriptR_d(|\calA_{c_k}|)=|\calA_{c_k}|$, i.e., the number of defenders assigned to a swarm $\calA_{c_k}$ is equal to the number of attackers in $\calA_{c_k}$. 

\begin{algorithm}[h]
		\caption{Gathering formations for the defenders }
		\label{alg:gathering_formations_centralized}
		
		\SetKwFunction{timeOptimalTraj}{timeOptimalTraj}
		\SetKwFunction{assignDtoGMILP}{assignDtoGMILP}
		\KwIn{${{\mathbf{r}}}_{d}(0)$,  $\mathbf{x}_{a}(0)$, $D_c(0)$, $\{A_{c_k}(0)|k \in I_{ac}(0)\}$ }
		\For{$k=1:N_{ac}(0)$}{
		CoM of $\calA_{c_k}(0)$: $\mathbf{x}_{ac_k}(0)= \sum_{i\in A_{c_k}(0)} \frac{\mathbf{x}_{ai}(0)} {|\calA_{c_k}(0)|}$;
		$\mathbf{P}_{ac_k}$=\timeOptimalTraj($\mathbf{x}_{ac_k}(0)$);
		}
		\While{ $\Sigma_{T_{lead}}>\epsilon_{tol}$}{
		$\Sigma_{T_{lead}}=0; \gamma_{ac_k}^{<}=0; \gamma_{ac_k}^{>}=\Gamma_{ac_k}-\rho_{pa}; \bm{\xi}^g=[\;];$\\
		\For{$k=1:N_{ac}(0)$}{
		$\gamma_{ac_k}=\frac{\gamma_{ac_k}^{<}+\gamma_{ac_k}^{>}}{2}$;\\
		$\mathbf{r}_{df_k}(0) = \mathscr{P}_{ac_k}({\gamma_{ac_k}})$; \\
		$\bm{\xi}^g_{c_k} =\mathbf{p}_{k}^{line}(\mathbf{r}_{df_k}(0),\vartheta_{ac_k}({\gamma_{ac_k}})-\pi)$\\
		$\bm{\xi}^g \gets \{\bm{\xi}^g,\bm{\xi}^g_{c_k}\}$;\\
		}
		$[\beta_{c}(0), \calT]$=\assignDtoGMILP(${\mathbf{r}}_{dc}(0)$, $\bm{\xi}^g $);\\
		\For{$k=1:N_{ac}$}{
		$\Sigma_{T_{lead}}=\Sigma_{T_{lead}}+|\frac{\gamma_{ac_k}}{\bar{v}_a} - \calT_{k}-\Delta T_{dc_k}^g|;$\\
		\eIf{$\frac{\gamma_{ac_k}}{\bar{v}_a} - \calT_{k}-\Delta T_d^g$<0}
		{$\gamma_{ac_k}^{>}=\gamma_{ac_k}$;\\}
		{$\gamma_{ac_k}^{<}=\gamma_{ac_k}$;}
		}
		}
		\Return{$\bm{\xi}^g$, $\beta_{c}(0)$,  $\{\mathbf{r}_{df_1}(0), \mathbf{r}_{df_2}(0),.... \mathbf{r}_{df_{N_{ac}(0)}}(0)\}$}
	\end{algorithm}
	In Algorithm \ref{alg:gathering_formations_centralized}, $\timeOptimalTraj(\mathbf{x}_{ac_k}(0))$ function finds the time-optimal trajectory $\mathbf{P}_{ac_k}$ for an agent starting at $\mathbf{x}_{ac_k}(0)$ to reach the protected area. The trajectory $\mathbf{P}_{ac_k}$ is associated with mappings $\mathscr{P}_{ac_k}:[0,\Gamma_{ac_k}] \rightarrow \bR^2$ and $\vartheta_{ac_k}:$ $[0,\Gamma_{ac_k}]\rightarrow [0,2\pi]$. Here $\mathscr{P}_{ac_k}(\gamma_{ac_k})$ gives the Cartesian coordinates, and $\vartheta_{ac_k}(\gamma_{ac_k})$ gives the direction of the tangent to the path at the location reached after traveling $\gamma_{ac_k}$ distance along the path from the initial position. ${\mathbf{r}}_{d}(0)=\{{\mathbf{r}}_{dj}(0)| j \in I_d\}$ is the set of initial positions of the defenders and $\mathbf{x}_{a}(0)=\{\mathbf{x}_{ai}(0)|i \in I_a\}$ is the set of initial states of the attackers. Each defender is assumed to have zero initial velocity\footnote{This is not a conservative assumption because if a defender has non-zero speed, one can apply acceleration opposite to its velocity to make the speed zero and assume the initial position for that defender to be the position at which this speed will become zero.}. The function \assignDtoGMILP assigns each defender $\calD_j$ in $\calD_{c}(0)$ initially located at ${\mathbf{r}}_{dj}(0)$ to one of the gathering locations in $\bm{\xi}^g=\{\bm{\xi}_{c_1}^g,\bm{\xi}_{c_2}^g,...,\bm{\xi}_{c_{N_{ac}(0)}}^g\}$ by solving the following the mixed integer linear program (MILP):
	\bse \label{eq:defender_gatheringPositions_assign_MILP}
	\begin{align}
	 \argmin_{\bm{\delta}} &  \ds \sum_{k = 1}^{N_{ac}(0)} \sum_{l = 1}^{|I_{dc_k}(0)|} \ds \sum_{j \in D_c(0)} \norm{{\mathbf{r}}_{dj}(0)-\bm{\xi}_{c_k,l}^g}\delta_{jl}^{c_k}  \label{eq:MILP_cost}\\
	\vspace{2mm}
	\text{Subject to } & \scriptstyle \sum_{k \in I_{ac}(0)} \sum_{l \in I_{dc_k}(0)} \delta_{jl}^{c_k}=1, \quad \forall j \in D_c(0); \label{eq:MILP_constraint_1}\\
	\vspace{2mm}
    & \scriptstyle \sum_{j \in D_c(0)} \delta_{jl}^{c_k}=1, \quad \forall l \in I_{dc_k}(0),\; \forall k \in I_{ac}(0), ; \label{eq:MILP_constraint_2}\\
	& \scriptstyle \delta_{jl}^{c_k}\in \{0,1\}, \quad \forall j \in D_c(0), \; \forall l \in I_{dc_k}(0),\; \forall k \in I_{ac}(0); \label{eq:MILP_constraint_3}
	\end{align}
	\ese
where the distance between an initial position ${\mathbf{r}}_{dj}(0)$ and $\bm{\xi}_{c_k,l}^g$ is used as the metric for solving the assignment problem, {the constraints \eqref{eq:MILP_constraint_1} ensure that each defender is assigned to a single goal location, the constraints \eqref{eq:MILP_constraint_2} ensure that each goal location is assigned a unique defender, and the last constraints \eqref{eq:MILP_constraint_3} force the decision variable $\delta_{jl}^{c_k}$ to be binary.} The decision variable $\delta_{jl}^{c_k}$ is 1 if the defender $\calD_j$ is assigned to go to the goal location $\bm{\xi}_{c_k,l}^g$ and 0 otherwise; and $\bm{\delta} \in \{0,1\}^{N_{\delta}(0) }$ is the binary decision vector defined as $\bm{\delta}=[\delta_{jl}^{c_k}|\forall j \in D_c(0), \; \forall l \in I_{dc_k}(0),\; \forall k \in I_{ac}(0)]^T$, where $N_{\delta}(0)=(N_d-|A_{uc}(0)|)\sum_{k \in I_{ac}(0)} \scriptR_d(|A_{c_k}(0)|)$. The function \assignDtoGMILP  also outputs $\calT=\{\calT_1,\calT_2,...,\calT_{N_{ac}}\}$, where $\calT_k$, for all $k \in I_{ac}(0)$, is the time required by the sub-team $\calD_{c_k}(0)$ to gather at their desired gathering formation. The parameter $\epsilon_{tol}>0$ is a user defined small number used as the convergence tolerance.

The idea in Algorithm \ref{alg:gathering_formations_centralized} is to find the gathering formations that are as far from the protected area as possible and each subteam $\calD_{c_k}(0)$ of the defenders is able to reach their assigned gathering formation at least $\Delta T_{dc_k}^g$ s before the center of mass (CoM) of $\calA_{c_k}$, that follows its time-optimal trajectory towards the protected area, reaches the center of the gathering formation. Here $\Delta T_{dc_k}^g$, for all $k \in I_{ac}(0)$ is a user-defined time that accounts for the size of the swarm $\calA_{c_k}$ and the time required to get connected by strings once arrived at the desired formation.

The Defender-to-Attacker-Swarm Assignment $\beta_{c_k}(0,\cdot)$, for all $k \in I_{ac}(0)$, is then obtained as:

\be
\beta_{c_k}(0, l ) = \argmax_{j} \delta_{jl}^{c_k*}
\ee
where $\delta_{jl}^{c_k*}$ is the optimal value of $\delta_{jl}^{c_k}$ from \eqref{eq:defender_gatheringPositions_assign_MILP}.

\subsection{Optimal assignment when attackers split at $t>0$}
In reaction to the defenders' attempt to herd the attackers, the attackers may split into new smaller swarms and/or scatter as individual attackers. We continuously track the radii of the clusters and run the DBSCAN algorithm only when at some instant $t=t_{se}$ the connectivity constraint is violated by the swarms of attackers $\calA_{c_k}(t_{se})$ for some $k \in I_{ac}(t_{se})$  i.e., when the radius $\rho_{ac_k}(t_{se})$ of the swarm of attackers $\calA_{c_k}(t_{se})$ exceeds the value $\bar{\rho}_{ac_k}(t_{se})=\frac{\bar{R}_{sb}}{2}\cot\left(\frac{\pi}{\scriptR_d(N_a)}\right) \frac{|\calA_{c_k}(t_{se})|-1}{N_a-1}$. The connectivity constraint violation is termed as split event in this paper. The split event is formally defined as:
	\begin{definition}[Split event] An instant $t_{se}$ when for any swarm $\calA_{c_k}(t_{se})$, $k \in I_{ac}(t_{se})$, the radius of the swarm of attackers $\calA_{c_k}(t_{se})$ defined as $\rho_{ac_k}(t_{se})=\max_{i \in A_{c_k}(t_{se})} \norm{\mathbf{r}_{ai}(t_{se})-\mathbf{r}_{ac_k}(t_{se})}$ exceeds the value $\bar{\rho}_{ac_k}(t_{se})$.
	\end{definition}
	
	We also make the following assumption regarding the splitting behavior of the attackers.
    \begin{assumption}
    Once a swarm of attackers splits, its member attackers never rejoin each other, i.e., for all $i 
\in I_a$, if $\exists \; t>0$ such that $\calA_{i} \notin \calA_{c_k}(t)$ for any $k \in I_{ac}(t)$ then $\calA_{i} \notin \calA_{c_k}(t')$ for all $t \le t'$. 
    \end{assumption}
The splitting behavior of the attackers requires re-assignment of the defenders, that were supposed to herd the given swarm of the attackers that just split, to the newly available interception or herding tasks. Next, we describe a mixed-integer quadratically constrained quadratic program (MIQCQP) to solve this assignment problem.
\subsubsection{{Decentralized} optimal assignment using MIQCQP}  When a swarm of attackers $\calA_{c_k}$ splits into smaller swarms at $t = t_{se}$. The newly identified swarms of the attackers by the DBSCAN algorithm are assigned new indices. Namely, one of the swarm is assigned the index $k$, i.e. the index of the parent swarm $\calA_{c_k}$ and the rest swarms are assigned integers greater than $N_{ac}(t_{se}^-)$ as their indices, where $t_{se}^-$ denotes the instant immediately before $t=t_{se}$.
  Let $A_{c}^{(k)}(t_{se})$ denote the indices of the clusters of the attackers that are newly formed out of the parent cluster $\calA_{c_k}(t_{se}^-)$, when the cluster $\calA_{c_k}$ splits at $t=t_{se}$, as identified by the DBSCAN algorithm. $\calA_{uc}^{(k)} (t_{se})$ is the set of unclustered attackers separated from the original cluster $\calA_{c_k}(t_{se})$ after the original cluster has split.
We aim to assign the defenders in $\calD_{c_k}(t_{se}^-)$, that are already connected via Open-StringNet $\calG_{sn}^{op}(D_{c_k}(t_{se}^-))$ and were tasked to herd the original cluster $\calA_{c_k}(t_{se}^-)$, to either intercept the unclustered attackers separated from the original cluster $\calA_{c_k}(t_{se}^-)$ or herd the smaller clusters formed by the attackers in the original swarm $\calA_{c_k}(t_{se}^-)$ after splitting. Herding the smaller swarms of the attackers still requires the sub-teams of the defenders to stay connected via Open-StringNets while the defenders assigned to intercept the unclustered attackers will now disconnect themselves from the rest of the Open-StrigNet. In \cite{chipade2020multiswarm}, we solved  a connectivity constrained generalized assignment problem (C2GAP) to assign connected sub-teams of the defenders to herd the newly formed sub-swarms of the attackers after the original attacking swarm splits. In contrast to that, the current assignment problem is more complex due to the requirement of assigning some individual defenders, who shall disconnect themselves from the rest of the Open-StringNet, to intercept the unclusterd attackers. 

Let $\delta_{jk'}^{herd}(t_{se})$ be the binary decision variable at time $t=t_{se}$ that takes value 1 if the defender $\calD_j$ is assigned to herd the swarm $\calA_{c_{k'}}(t_{se})$ and 0 otherwise. We formulate the MIQCQP in \eqref{eq:defenders_to_attackers_and_swarm_assign_MIQCQP} to assign the defenders on the Open-StringNet $\calG_{sn}^{op}(D_{c_k}(t_{se}^-))$ to herd the newly formed swarms of the attackers, $\calA_{c_{k'}}(t_{se})$, for all $k' \in A_{c}^{(k)}(t_{se})$, and the unclustered attackers $\calA_{uc}^{(k)}(t_{se})$.
\setlength{\abovedisplayskip}{10pt}
\begin{figure*}
	\bse \label{eq:defenders_to_attackers_and_swarm_assign_MIQCQP}
	\begin{align}
	\bm{\delta}^{(k)*}(t_{se}) = \argmin_{\bm{\delta}^{(k)}(t_{se})}  & \displaystyle \sum_{k' \in A_{c}^{(k)}(t_{se})} \sum_{j\in D_{c_k}(t_{se}^-)} \norm{\mathbf{r}_{{ac}_{k'}}(t_{se})-\mathbf{r}_{dj}(t_{se})}\delta_{jk'}^{herd}(t_{se}) +\sum_{i \in A_{uc}^{(k)}(t_{se})} \sum_{j\in D_{c_k}(t_{se}^-)} C_d^{int}(\mathbf{X}_{dj}^{ai})\delta_{ji}^{int}(t_{se}) \nonumber\\
	& + \sum_{i, i' \in A_{uc}^{(k)}(t_{se})} \sum_{j, j'\in D_{c_k}(t_{se}^-)} C_d^{col}(\mathbf{X}_{dj}^{ai}, \mathbf{X}_{dj^{\prime}}^{ai^{\prime}})\delta_{ji}^{int}(t_{se}) \delta_{j'i'}^{int}(t_{se})  \label{eq:MIQCQP_cost}\\
	\text{Subject to } & \textstyle \sum_{k' \in A_{c}^{(k)}(t_{se})} \delta_{jk'}^{herd}(t_{se}) + \sum_{i \in A_{uc}^{(k)}(t_{se})} \delta_{ji}^{int}(t_{se}) = 1, \quad \forall j \in D_{c_k}(t_{se}^-); \label{eq:MIQCQP_constraint_1}\\
	\vspace{2mm}
	\vspace{2mm}
    & \textstyle \sum_{j \in D_{c_k}(t_{se}^-)} \delta_{jk'}^{herd}(t_{se}) = \scriptR_d(|\calA_{c_{k'}}(t_{se})|), \quad \forall k' \in A_{c}^{(k)}(t_{se}); \label{eq:MIQCQP_constraint_3}\\
    	\vspace{2mm}
    &\textstyle \sum_{j \in D_{c_k}(t_{se}^-)} \delta_{ji}^{int} (t_{se}) =1, \quad \forall i \in A_{uc}^{(k)}(t_{se}); \label{eq:MIQCQP_constraint_4}\\
	\vspace{2mm}
	&\textstyle \sum_{l \in I_{dc_k}'} \delta_{\beta^-_k(l)k'}^{herd}(t_{se}) \delta_{\beta^-_k(l+1)k'}^{herd}(t_{se}) \ge \scriptR_d(|\calA_{c_{k'}}(t_{se})|)-1, \quad \forall k' \in A_{c}^{(k)}(t_{se}); \label{eq:MIQCQP_constraint_5}\\
	& \textstyle \sum_{j \in D_{c_k}(t_{se}^-)} \Bigl (\sum_{k' \in A_{c}^{(k)}(t_{se})} \delta_{jk'}^{herd}(t_{se}) + \sum_{i \in A_{uc}^{(k)}(t_{se})}  \delta_{ji}^{int}(t_{se}) \Bigr ) =|D_{c_k}(t_{se}^-)|; \label{eq:MIQCQP_constraint_6}\\
	& \textstyle \delta_{jk'}^{herd}(t_{se}), \; \delta_{ji}^{int}(t_{se})\in \{0,1\}, \quad \forall j \in D_{c_k}(t_{se}^-), \; k' \in A_{c}^{(k)}(t_{se}), \; i \in A_{uc}^{(k)}(t_{se});
	\end{align}
	\ese
	\end{figure*}
In \eqref{eq:defenders_to_attackers_and_swarm_assign_MIQCQP}, $\bm{\delta}^{(k)}(t_{se}) \in \{0,1\}^{N_{\bm{\delta}^{(k)}}(t_{se})}$ is the binary decision vector defined as $\bm{\delta}^{(k)}(t_{se})=[[\delta_{jk'}^{herd}(t_{se})| k' \in A_{c}^{(k)}(t_{se}), j \in D_{c_k}(t_{se}^-)],[\delta_{ji}^{int}(t_{se})| i \in A_{uc}^{(k)}(t_{se}), j \in D_{c_k}(t_{se}^-)]]^T$, where $N_{\bm{\delta}^{(k)}}(t_{se}) = |\calD_{c_k}(t_{se}^-)|\left(|A_{c}^{(k)}(t_{se})|+|A_{uc}^{(k)}(t_{se})|\right)$; $I_{dc_k}'=\{1,2,...,|\calD_{c_k}|-1\}$; and $\beta^-_k(l) = \beta_{c_k}(t_{se}^-,l)$.

The optimization cost in \eqref{eq:defenders_to_attackers_and_swarm_assign_MIQCQP} is the sum of distances of the defenders from the centers of the attackers' swarms to which they are assigned, the times to capture required by the defenders to capture the unclustered attacker that are assigned to them, and the collision costs incurred by the defenders that are assigned interception task. This ensures that the collective effort needed by all the defenders is minimized when enclosing the swarms of the attackers and that the unclustered attackers are captured as quickly as possible while minimizing any possible collisions among the fast moving defenders that are assigned the interception task. The constraints \eqref{eq:MIQCQP_constraint_1} ensure that each of the defenders in $\calD_{c_k}(t_{se}^-)$ is assigned either to exactly one unclustered attacker or to exactly one swarm of the attackers. 
The  capacity constraints \eqref{eq:MIQCQP_constraint_3} ensure that for all $k' \in A_{c}^{(k)}(t_{se})$, the swarm $\calA_{c_{k'}}(t_{se})$ has exactly $\scriptR_d(|\calA_{c_{k'}}(t_{se})|)$ defenders assigned to it. The constraints \eqref{eq:MIQCQP_constraint_4} ensure that each unclustered attacker in $\calA_{uc}^{(k)}(t_{se})$ has exactly one of the terminal defenders assigned to it. The quadratic constraints \eqref{eq:MIQCQP_constraint_5} ensure that all the defenders assigned to swarm $\calA_{c_{k'}}(t_{se})$ are connected together with an underlying Open-StringNet for all $k' \in A_{c}^{(k)}$ and the constraint \eqref{eq:MIQCQP_constraint_6} ensures that all the $|D_{c_k}(t_{se}^-)|$ defenders are assigned to the attackers' swarms and the unclustered attackers. 
	
    {The aforementioned MIQCQP \eqref{eq:defenders_to_attackers_and_swarm_assign_MIQCQP} is solved by the lead defender in $\calD_{c_k}(t_{se}^-)$, where the lead defender is identified to be the one in the middle of the Open-StringNet formation, i.e., the defender  $\calD_{\beta_k(t_{se}^-,l_{i}}) $ where $l_i= \floor{\frac{|\calD_{c_k}(t_{se}^-)|}{2}}$, for all $k$ for which the $\calA_{c_k}$ have split. 
    This helps the defenders find the Defender-to-Attacker-Swarm
    assignment quickly, and without having to consider all the  agents
    in the assignment formulation, i.e., in a decentralized way. }
    
    The aforementioned MIQCQP \eqref{eq:defenders_to_attackers_and_swarm_assign_MIQCQP} can be solved using a MIP solver Gurobi \cite{gurobi}. After solving \eqref{eq:defenders_to_attackers_and_swarm_assign_MIQCQP}, one can find the mapping $\beta_{c_{k'}}(t,\cdot)$, for all $k' \in A_{c}^{(k)}(t_{se})$, as follows:
\be
\beta_{c_{k'}}(t, l)= \beta_{c_k}^- (l_0+l), \quad \text{     $\forall t \in [t_{se}+t_{comp}, \; t_{se}^{next} ]$},
\ee
where $l_0$ is the smallest integer for which $\delta_{\beta_{c_k}^-(l_0+1)k}  (t_{se}) = 1$; $t_{comp}$ is the computation time to solve \eqref{eq:defenders_to_attackers_and_swarm_assign_MIQCQP}; and $t_{se}^{next}$ is an unknown future time at which a split happens. In other words, the assignment obtained using the states at $t_{se}$ continues to be a valid assignment until the next split event happens at some unknown time $t_{se}^{next}$ in the future.
The worst-case time complexity of the MIQCQP in \eqref{eq:defenders_to_attackers_and_swarm_assign_MIQCQP} is: 
\be
C_{M}^{comp}(t_{se},k) = O(2^{N_{\bm{\delta}^{(k)}}(t_{se})})
\ee
where $N_{\bm{\delta}^{(k)}}(t_{se}) = |\calD_{c_k}(t_{se}^-)|\left(|A_{c}^{(k)}(t_{se})|+|A_{uc}^{(k)}(t_{se})|\right)$.
	 
\subsection{Suboptimal assignment when attackers split at $t>0$}
\subsubsection{Assignment using reduced-size MIQCQP (rs-MIQCQP)}  The worst-case complexity $C_{M}^{comp}(t_{se},k)$ of the MIQCQP in \eqref{eq:defenders_to_attackers_and_swarm_assign_MIQCQP} can be reduced further under certain assumption on the behavior of the attackers. Let us first define a conical envelope around the center of a swarm.

\begin{definition}[Conical Envelope]
A conical envelope $E_{con}(\mathbf{r}_0,\psi)$, centered at $\mathbf{r}_0=[x_0,y_0]^T$ is defined as
$E_{con}(\mathbf{r}_0, \psi) = \bigl\{\{(x,y)\in \bR^2|y-y_0-m_1(x-x_0)>0\}\cap\{(x,y)\in \bR^2|y-y_0-m_2(x-x_0)<0\}\bigr\} \cup \bigl\{\{(x,y)\in \bR^2|y-y_0-m_1(x-x_0)<0\}\cap\{(x,y)\in \bR^2|y-y_0-m_2(x-x_0)>0\} \bigr\}$, where $m_1 = \tan \left(\tan^{-1}(\frac{y_0-y_p}{x_0-x_p}) -\frac{\pi}{2} -\psi \right)$ and $m_2 = \tan \left(\tan^{-1}(\frac{y_0-y_p}{x_0-x_p}) -\frac{\pi}{2} +\psi \right)$.

\end{definition}

\begin{figure}
	\centering
	\includegraphics[width=.9\linewidth,trim={10.8cm 1.2cm 8.3cm .0cm},clip]{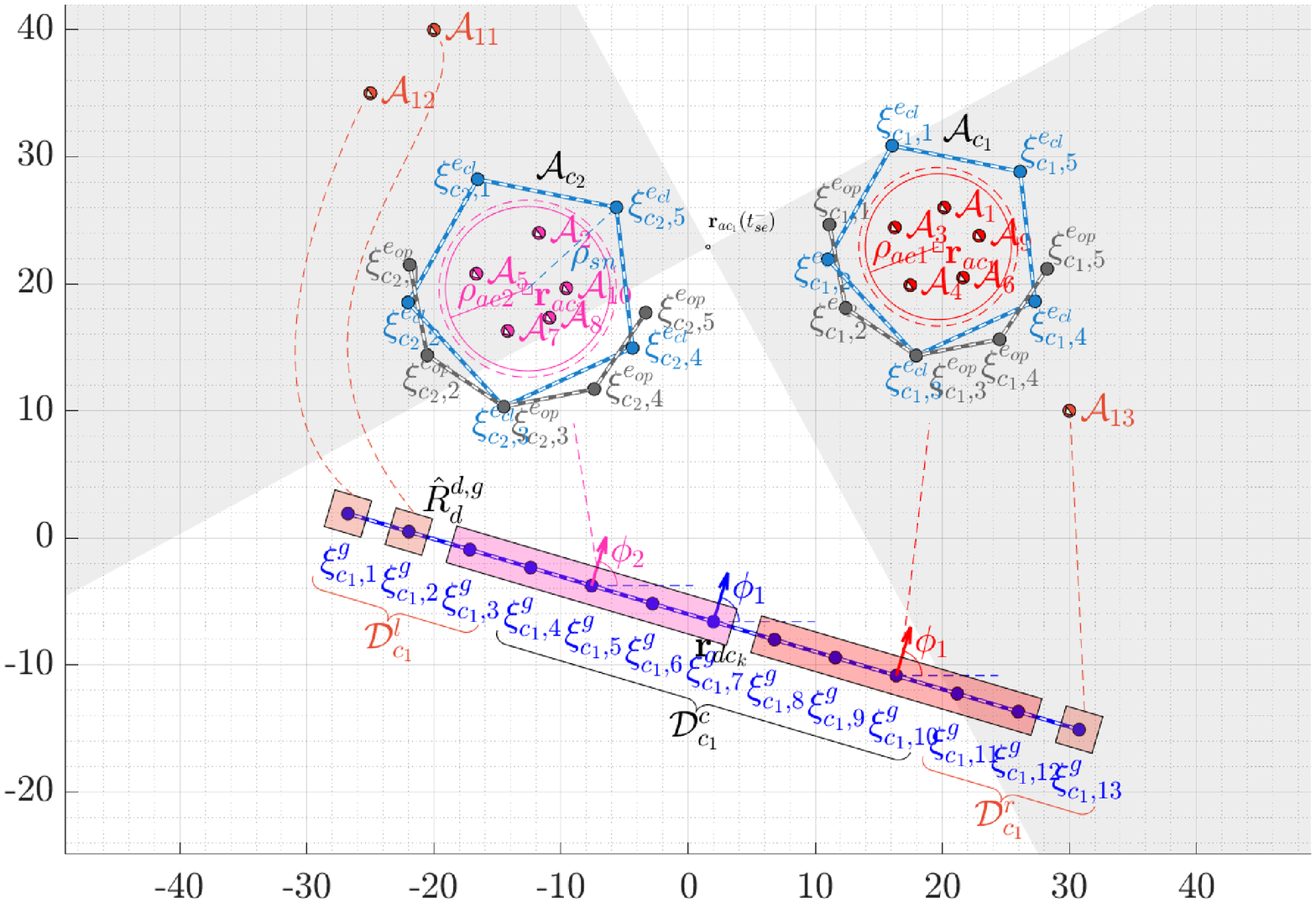}
	\caption{Assignment of the defenders after the attackers split using rs-MIQCQP}
	\label{fig:assignmentAfterSplit}
    \end{figure}

\begin{assumption}\label{assum:attackers_split}
A swarm of the attackers $\calA_{c_k}$, for any $k$, splits at $t=t_{se}$, such that all the unclustered attackers (swarms with less than 3 attackers) are the farthest from the center of the original swarm $\calA_{c_k}(t_{se}^-)$ and their centers lie within the conical envelope $E_{con}(\mathbf{r}_{ac_k}(t_{se}^-),\frac{\pi}{4})$, i.e., $\forall i \in A_{uc}^{(k)}(t_{se})$, $\norm{\mathbf{r}_{ai}(t_{se}) - \mathbf{r}_{ac_k}(t_{se}^-)} >  \max_{k' \in A_{c}^{(k)}(t_{se})} \norm{\mathbf{r}_{ac_{k'}}(t_{se}) - \mathbf{r}_{ac_k}(t_{se}^-)}$ and $\mathbf{r}_{ai}(t_{se}) \in E_{con}(\mathbf{r}_{ac_k}(t_{se}^-),\frac{\pi}{4})$ (gray shaded region in Fig.~\ref{fig:assignmentAfterSplit}).
\end{assumption}
Assumption \ref{assum:attackers_split} implies that the unclustered attackers aim to spread in the direction transverse to the direction toward the protected area because of the presence of the defenders in front of them in order to maximize their chances of not getting captured by the defenders and reaching the protected area. 
Under Assumption \ref{assum:attackers_split}, we can assign only the defenders from either end of the Open-StringNet to intercept the unclustered attackers while assign the defenders in the central part of the Open-StringNet to herd the newly formed clusters of the attackers. 

 Let $\calD_{c_k}^l(t_{se}^-)=\{\calD_j|j \in D_{c_k}^l(t_{se}^-)\}$ be the group of $|A_{uc}^{(k)}(t_{se})|$ defenders at the left end of the Open-StringNet $\calG_{sn}^{op}(D_{c_k}(t_{se}^-))$, where $D_{c_k}^l(t_{se}^-)=\{\beta_{c_k}^-(1),\beta_{c_k}^-(2),...,\beta_{c_k}^-(|A_{uc}^{(k)}(t_{se})|)\}$. Here the left end of the Open-StringNet formation refers to the end approached first when one rotates anti-clockwise standing at the center $\mathbf{r}_{df_k}$ and starting when facing in the direction $\phi_k$ of the formation (see Fig.~\ref{fig:assignmentAfterSplit}). Similarly, let $\calD_{c_k}^r(t_{se}^-)=\{\calD_j|j \in D_{c_k}^r(t_{se}^-)\}$ be the group of $|A_{uc}^{(k)}(t_{se})|$ defenders at the right end of the Open-StringNet formation $\calG_{sn}^{op}(D_{c_k}(t_{se}^-))$, where $D_{c_k}^r(t_{se}^-)=\{\beta_{c_k}^-(|D_{c_k}(t_{se}^-)|-|A_{uc}^{(k)}|+1),\beta_{c_k}^-(|D_{c_k}(t_{se}^-)|-|A_{uc}^{(k)}|+2),...,\beta_{c_k}^-(|D_{c_k}(t_{se}^-)|)\}$ (see Fig.~\ref{fig:assignmentAfterSplit}). Let us call $\calD_{c_k}^t(t_{se}^-)=\{\calD_j | j \in D_{c_k}^l(t_{se}^-) \cup D_{c_k}^r(t_{se}^-)\}$ as the group of terminal defenders of the Open-StringNet $\calG_{sn}^{op}(D_{c_k}(t_{se}^-))$. We denote by $\calD_{c_k}^c(t_{se}^-)=\{\calD_j|j \in D_{c_k}^c(t_{se}^-)\}$ the central defenders, the group of the defenders excluding the terminal defenders $\calD_{c_k}^t(t_{se}^-)$, where $D_{c_k}^c(t_{se}^-)= D_{c_k}(t_{se}^-)\backslash D_{c_k}^t(t_{se}^-)$.
 
 Next, we develop a reduced-size MIQCQP, in which only the terminal defenders $\calD_{c_k}^t(t_{se}^-)$ are assigned the interception task, in \eqref{eq:defenders_to_attackers_and_swarm_assign_MIQCQP2}.
 \begin{figure*}
	\bse \label{eq:defenders_to_attackers_and_swarm_assign_MIQCQP2}
	\begin{align}
	\bm{\delta}_{rs}^{(k)*}(t_{se}) = \argmin_{\bm{\delta}_{rs}^{(k)}(t_{se})}  & \displaystyle \sum_{k' \in A_{c}^{(k)}(t_{se})} \sum_{j\in D_{c_k}(t_{se}^-)} \norm{\mathbf{r}_{{ac}_{k'}}(t_{se})-\mathbf{r}_{dj}(t_{se})}\delta_{jk'}^{herd}(t_{se}) +\sum_{i \in A_{uc}^{(k)}(t_{se})} \sum_{j\in D_{c_k}^t(t_{se}^-)} C_d^{int}(\mathbf{X}_{dj}^{ai})\delta_{ji}^{int}(t_{se}) \nonumber\\
	& + \sum_{i, i' \in A_{uc}^{(k)}(t_{se})} \sum_{j, j'\in D_{c_k}^t(t_{se}^-)} C_d^{col}(\mathbf{X}_{dj}^{ai}, \mathbf{X}_{dj^{\prime}}^{ai^{\prime}})\delta_{ji}^{int}(t_{se}) \delta_{j'i'}^{int}(t_{se})  \label{eq:MIQCQP2_cost}\\
	\text{Subject to } & \textstyle \sum_{k' \in A_{c}^{(k)}(t_{se})} \delta_{jk'}^{herd}(t_{se}) + \sum_{i \in A_{uc}^{(k)}(t_{se})} \delta_{ji}^{int}(t_{se}) = 1, \quad \forall j \in D_{c_k}^t(t_{se}^-); \label{eq:MIQCQP2_constraint_1}\\
	\vspace{2mm}
	& \textstyle \sum_{k' \in A_{c}^{(k)}(t_{se})} \delta_{jk'}^{herd}(t_{se}) =1, \quad \forall j \in D_{c_k}^c(t_{se}); \label{eq:MIQCQP2_constraint_2}\\
	\vspace{2mm}
    & \textstyle \sum_{j \in D_{c_k}(t_{se}^-)} \delta_{jk'}^{herd}(t_{se}) = \scriptR_d(|\calA_{c_{k'}}(t_{se})|), \quad \forall k' \in A_{c}^{(k)}(t_{se}); \label{eq:MIQCQP2_constraint_3}\\
    	\vspace{2mm}
    &\textstyle \sum_{j \in D_{c_k}^t(t_{se}^-)} \delta_{ji}^{int} (t_{se}) =1, \quad \forall i \in A_{uc}^{(k)}(t_{se}); \label{eq:MIQCQP2_constraint_4}\\
	\vspace{2mm}
	&\textstyle \sum_{l \in I_{dc_k}'} \delta_{\beta^-_k(l)k'}^{herd}(t_{se}) \delta_{\beta^-_k(l+1)k'}^{herd}(t_{se}) \ge \scriptR_d(|\calA_{c_{k'}}(t_{se})|)-1, \quad \forall k' \in A_{c}^{(k)}(t_{se}); \label{eq:MIQCQP2_constraint_5}\\
	& \textstyle \sum_{j \in D_{c_k}^c(t_{se}^-)} \sum_{k' \in A_{c}^{(k)}(t_{se})} \delta_{jk'}^{herd}(t_{se}) + \sum_{j \in D_{c_k}^t(t_{se}^-)} \sum_{i \in A_{uc}^{(k)}(t_{se})}  \delta_{ji}^{int}(t_{se})  =|D_{c_k}(t_{se}^-)|; \label{eq:MIQCQP2_constraint_6}\\
	& \textstyle \delta_{jk'}^{herd}(t_{se}), \; \delta_{ji}^{int}(t_{se})\in \{0,1\}, \quad \forall j \in D_{c_k}(t_{se}^-), \; k' \in A_{c}^{(k)}(t_{se}), \; i \in A_{uc}^{(k)}(t_{se});
	\end{align}
	\ese
	\end{figure*}
In \eqref{eq:defenders_to_attackers_and_swarm_assign_MIQCQP2} the length of the decision vector $\bm{\delta}_{rs}^{(k)*}(t_{se}) = [[\delta_{jk'}^{herd}(t_{se})| k' \in A_{c}^{(k)}(t_{se}), \; j \in D_{c_k}(t_{se}^-)],[\delta_{ji}^{int}(t_{se})| i \in A_{uc}^{(k)}(t_{se}), \;j \in D_{c_k}(t_{se}^-)]]^T$ is $N_{\bm{\delta}_{rs}^{(k)}}(t_{se}) = |\calD_{c_k}(t_{se}^-)||A_{c}^{(k)}(t_{se})|+\min(2|A_{uc}^{(k)}(t_{se})|, |\calD_{c_k}(t_{se}^-)|) |A_{uc}^{(k)}(t_{se})| $. We have the following result about the computation cost of \eqref{eq:defenders_to_attackers_and_swarm_assign_MIQCQP2}.
 
 \begin{lemma} \label{lem:computation_cost_MIQCQP2}
The worst-case computational cost of \eqref{eq:defenders_to_attackers_and_swarm_assign_MIQCQP2}, $C_{rsM}^{comp}(t_{se},k)$, satisfies:
\be
C_{rsM}^{comp}(t_{se},k) = O(2^{N_{\bm{\delta}_{rs}^{(k)}}(t_{se})}) \le C_{M}^{comp}(t_{se},k).
\ee 
Furthermore, if the number of unclustered attackers is less than half of the total number of attackers in the original cluster, i.e., $|A_{uc}^{(k)}(t_{se})|<\frac{|A_{c_k}(t_{se}^-)|}{2}$, then $C_{rsM}^{comp}(t_{se},k) < C_{M}^{comp}(t_{se},k)$.
 \end{lemma}

Figure~\ref{fig:assignmentAfterSplit} shows an instance of the assignment of the defenders on the Open-StringNet $\calG_{sn}^{op}(D_{c_1}(t_{se}^-))$ at some time $t=t_{se}$, where $D_{c_1}(t_{se}^-)=\{1,2,3,...,13\}$, to the newly formed clusters $\calA_{c_1}(t_{se})=\{\calA_{1},\calA_{3},\calA_{4},\calA_{6},\calA_{9}\}$, $\calA_{c_2}(t_{se})=\{\calA_{2},\calA_{5},\calA_{7},\calA_{8},\calA_{10}\}$ and the unclustrered attackers $\calA_{uc}^{(1)}(t_{se})=\{\calA_{11},\calA_{12},\calA_{13}\}$.
	 After solving the rs-MIQCQP \eqref{eq:defenders_to_attackers_and_swarm_assign_MIQCQP2}, as shown in Fig.~\ref{fig:assignmentAfterSplit}, defenders $\calD_{\beta_{1}^{-}(1)}$, $\calD_{\beta_{1}^{-}(2)}$ and $\calD_{\beta_{1}^{-}(13)}$ are assigned to the unclustered attackers $\calA_{12}$, $\calA_{11}$, $\calA_{13}$, respectively, so that these attackers can be intercepted as soon as possible. The connected sub-teams $\{\calD_{\beta_{1}^{-}(8)},\calD_{\beta_{1}^{-}(9)},\calD_{\beta_{1}^{-}(10)},\calD_{\beta_{1}^{-}(11)},\calD_{\beta_{1}^{-}(12)}\}$ and $\{\calD_{\beta_{1}^{-}(3)},\calD_{\beta_{1}^{-}(4)},\calD_{\beta_{1}^{-}(5)},\calD_{\beta_{1}^{-}(6)},\calD_{\beta_{1}^{-}(7)}\}$ are assigned to the newly formed swarms of the attackers $\calA_{c_1}(t_{se})$ and $\calA_{c_2}(t_{se})$, respectively.
	 
\subsubsection{Hierarchical approach to assignment (a heuristic)}
Finding the optimal assignment of the defenders for interception and herding tasks by solving the MIQCQPs \eqref{eq:defenders_to_attackers_and_swarm_assign_MIQCQP} and \eqref{eq:defenders_to_attackers_and_swarm_assign_MIQCQP2} may not be real-time implementable for a large number of agents $(>100)$. In this subsection, we develop a computationally efficient hierarchical approach to find the defender-to-attacker-swarm assignment under Assumption \ref{assum:attackers_split}. The idea is to split a large dimensional assignment problem into smaller, low-dimensional assignment problems that can be solved optimally and quickly.

Let $\scriptA_k(t_{se})$ be a data structure that stores information about the attackers in $\calA_{c_k}(t_{se}^-)$ and has data fields: $\scriptA_k(t_{se}).\mathbf{r}_{ac}=[\mathbf{r}_{{ac}_{k'}}| k' \in A_{c}^{(k)}(t_{se})]$, centers of the newly formed attackers' swarms after separating from the original swarm $\calA_{c_k}(t_{se}^-)$; $\scriptA_k(t_{se}).\mathbf{n}_{ac}=[|\calA_{c_{k'}}(t_{se})|| k' \in A_{c}^{(k)}(t_{se})]$, numbers of the attackers in each swarm; $\scriptA_k(t_{se}).N_{ac} = |A_{c}^{(k)}(t_{se})|$, total number of attackers' clusters formed from $\calA_k(t_{se}^{-})$; $\scriptA_k(t_{se}).\mathbf{r}_{uc}=[\mathbf{r}_{{ai}}| i \in A_{uc}^{(k)}(t_{se})]$ current states of the unclustered attackers in $\calA_{uc}^{(k)}(t_{se})$;  $\scriptA_k(t_{se}).N_{uc}$, total number of unclustered attackers; $\scriptA_k(t_{se}).N_a=|\calA_{c_k}(t_{se}^-)|$, total number of attackers $\calA_{c_k}(t_{se}^-)$. Similarly, $\scriptD_k(t_{se})$ is a data structure that stores the information of the defenders on the original Open-StringNet $\calG_{sn}^{op}(D_{c_k}(t_{se}^-))$ with data fields: $\scriptD_k(t_{se}).\mathbf{r}_{d}=[\mathbf{r}_{dj}| j \in D_{c_k}(t_{se}^-)]$, positions of the defenders on $\calG_{sn}^{op}(D_{c_k}(t_{se}^-))$; and  $\scriptD_k(t_{se}).\beta=\beta_{c_k}(t_{se}^-)$, the original assignment mapping of the defenders on the Open-StringNet $\calG_{sn}^{op}(D_{c_k}(t_{se}^-))$.

Algorithm \ref{alg:defenders_to_attackers_and_swarm_assign_Hierarchical} provides the steps to solve the assignment problem quickly by hierarchically reducing the original big assignment problem into smaller ones.	 

\begin{figure}
	\centering
	\includegraphics[width=1\linewidth,trim={2.5cm 1.7cm 1cm .7cm},clip]{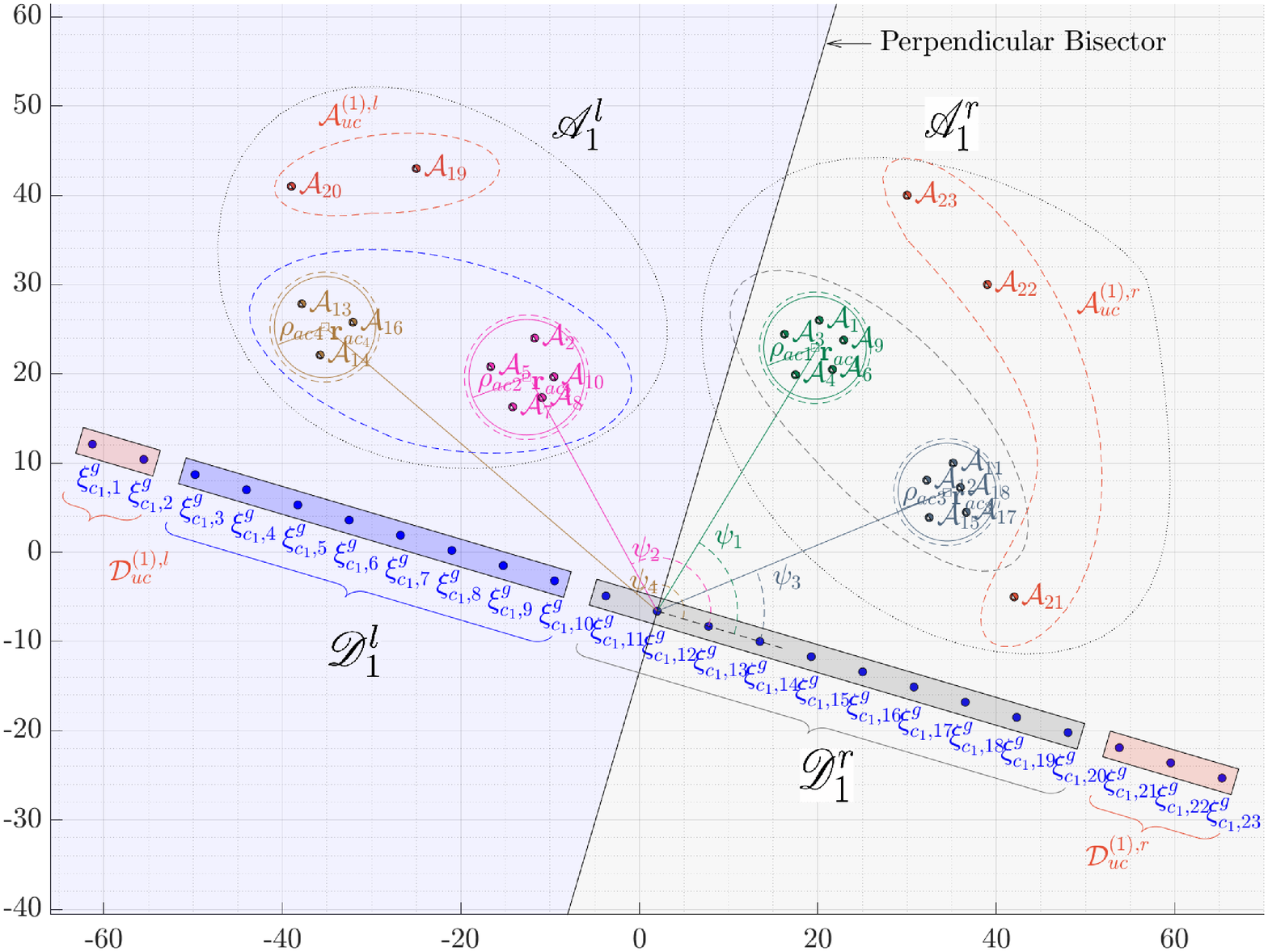}
	\caption{Grouping for the hierarchical algorithm}
	\label{fig:groupsAfterSplit}
    \end{figure}

\begin{algorithm}[h]
		\caption{Defender-to-Attacker-Swarm Assignment (DASA)}
		\label{alg:defenders_to_attackers_and_swarm_assign_Hierarchical}
		\KwIn{$\scriptA_k(t_{se}), \scriptD_k(t_{se})$}
		\SetKwFunction{assignHierarchical}{assignHierarchical}
		\SetKwFunction{assignMIQCQP}{assignMIQCQP}
		\SetKwFunction{splitClustersEqual}{splitClustersEqual}
		\SetKwFunction{splitUnclustAtt}{splitUnclustAtt}
		\SetKwFunction{CADAA}{CADAA}
		\SetKwProg{Fn}{Function}{:}{}
		[$\calA_{uc}^{(k),l}(t_{se}), \; \calD_{uc}^{(k),l}(t_{se}),\calA_{uc}^{(k),r}(t_{se}),\calD_{uc}^{(k),r}(t_{se})$] =\\
		\qquad \qquad \qquad \splitUnclustAtt($\scriptA_k(t_{se}),\scriptD_k(t_{se}))$;\\
		$\beta_{uc}^{(k),l}$= \CADAA($\calA_{uc}^{(k),l}(t_{se}),\calD_{uc}^{(k),l}(t_{se})$);\\
		$\beta_{uc}^{(k),r}$= \CADAA($\calA_{uc}^{(k),r}(t_{se}),\calD_{uc}^{(k),r}(t_{se})$);\\
		$\beta_{uc} (t_{se})\gets \{ \beta_{uc} (t_{se}),\; \beta_{uc}^{(k),l} \cup \beta_{uc}^{(k),r} \}$;\\
		$\scriptD_k(t_{se}).\calD_{c_k}\gets(\scriptD_k(t_{se}).\calD_{c_k})\backslash (\calD_{uc}^{(k),l}(t_{se}) \cup \calD_{uc}^{(k),r}(t_{se}));$\\
		$\beta_{c}(t_{se}) \gets \{ \beta_{c}(t_{se}), \; \assignHierarchical{$\scriptA_k(t_{se}), \scriptD_k(t_{se})$}\}$;\\
		\Return{$\beta_{uc}(t_{se}), \beta_{c}(t_{se})$};
		\vspace{2mm}
		\hrule
		\vspace{.5mm}
		\hrule
        \Fn{\assignHierarchical{$\scriptA_k, \scriptD_k$}}{
		\eIf{$\scriptA_k.N_{ac}>\underline{N}_{ac}$}{[$\scriptA_k^l,\scriptD_k^l,\scriptA_k^r,\scriptD_k^r$] =\\
		\quad \splitClustersEqual($\scriptA_k,\scriptD_k)$;\\
		\For{$\iota \in \{l,r\}$}{
		\eIf{$\scriptA_k^{\iota}.N_{ac}>\underline{N}_{ac}$}{
		$\beta_{c_k}^{\iota}=$ \assignHierarchical($\scriptA_k^{\iota}, \scriptD_k^{\iota}$);}
		{$\beta_{c_k}^{\iota}=$ \assignMIQCQP($\scriptA_k^{\iota},\scriptD_k^{\iota}$);}
		}
		$\beta_{c}=\{\beta_{c_k}^l,\beta_{c_k}^r\};$
		}
		{$\beta_{c}$=\assignMIQCQP($\scriptA_k(t_{se}),\scriptD_k$);}
		\Return{$\beta_{c}$};
		}
		\hrule
		\vspace{.5mm}
		\hrule
\end{algorithm}
	In Algorithm \ref{alg:defenders_to_attackers_and_swarm_assign_Hierarchical}, the function \splitUnclustAtt($\scriptA_k(t_{se}),\scriptD_k(t_{se}))$ splits the unclustered attackers $\calA_{uc}^{(k)}(t_{se})$ into two groups: left group $\calA_{uc}^{(k),l}(t_{se})$ and right group $\calA_{uc}^{(k),r}(t_{se})$. The normal bisector of the line segment joining the positions $\mathbf{r}_{d\beta_{c_k}^{-}(1)}(t_{se})$ and $\mathbf{r}_{d\beta_{c_k}^{-}(|D_{c_k}|)}(t_{se})$ acts as separating hyperplane for the groups $\calA_{uc}^{(k),l}(t_{se})$ and $\calA_{uc}^{(k),r}(t_{se})$. The unclustered attackers that lie in the half-plane containing the left side of Open-StringNet {and the normal bisector itself} are part of the left group $\calA_{uc}^{(k),l}(t_{se})$ and the rest unclustered attackers in $\calA_{uc}^{(k)}(t_{se})$ are part of the right group $\calA_{uc}^{(k),r}(t_{se})$ (see Fig.~\ref{fig:groupsAfterSplit}). The function \splitUnclustAtt also outputs $\calD_{uc}^{(k),l}(t_{se})$, the leftmost $|\calA_{uc}^{(k),l}(t_{se})|$ defenders on the Open-StringNet $\calG_{sn}^{op}(D_{c_k}(t_{se}^-))$; and $\calD_{uc}^{(k),r}(t_{se})$, the rightmost $|\calA_{uc}^{(k),r}(t_{se})|$ defenders on the Open-StringNet $\calG_{sn}^{op}(D_{c_k}(t_{se}^-))$ (see Fig.~\ref{fig:groupsAfterSplit}).
	The function \CADAA($\calA_{uc}^{(k),l}(t_{se}),\calD_{uc}^{(k),l}(t_{se})$) assigns the defenders in $\calD_{uc}^{(k),l}(t_{se})$ to intercept the attackers $\calA_{uc}^{(k),l}(t_{se})$ by solving CADAA \eqref{eq:defender_attackers_assign_MIQP}. Line 6 in Algorithm \ref{alg:defenders_to_attackers_and_swarm_assign_Hierarchical} removes the the defenders in  $\calD_{uc}^{(k),l}(t_{se})$ and $\calD_{uc}^{(k),r}(t_{se})$, that are already assigned to intercept the unclustered attackers, from further processing. The function \assignHierarchical{$\scriptA_k(t_{se}), \scriptD_k(t_{se})$} then assigns the remaining connected defenders on the Open-StringNet to the clusters of the attackers $\{\calA_{c_{k'}} (t_{se})| k' \in A_{c}^{(k)}(t_{se})\}$.




	
	In the function \assignHierarchical, the function \splitClustersEqual($\scriptA_k(t_{se}), \scriptD_k(t_{se})$) splits the clusters of the attackers into two groups $\scriptA_k^l(t_{se})$ and $\scriptA_k^r(t_{se})$ of roughly equal number of attackers and the defenders into two groups $\scriptD_k^l(t_{se})$ and $\scriptD_k^r(t_{se})$. The split is performed based on the angles $\psi_{k'}$ made by relative vectors $\mathbf{r}_{{ac}_{k'}}(t_{se})-\mathbf{r}_{dc_k}(t_{se}^-)$, for all $k' \in A_{c}^{(k)}(t_{se})$, with the vector $\mathbf{r}_{d\jmath_t}(t_{se}^-)-\mathbf{r}_{dc_k}(t_{se}^-)$ where $\mathbf{r}_{dc_k}(t_{se}^-)=\frac{\mathbf{r}_{d \jmath_1}(t_{se})+\mathbf{r}_{d\jmath_t}(t_{se})}{2}$ is the center of $\calD_{c_k}(t_{se}^-)$, where $\jmath_1 = \beta_{c_k}^-(1) $ and $\jmath_t = \beta_{c_k}^-(|D_{c_k}(t_{se}^-)|))$. We first arrange these angles $\psi_{k'}$ in the descending order. The first few clusters in the arranged list with roughly half the total number of attackers become the left group $\scriptA_k^l(t_{se})$ and the rest become the right group $\scriptA_k^r(t_{se})$ (see Fig.~\ref{fig:groupsAfterSplit}). Similarly, the left group $\scriptD_k^l(t_{se})$ is formed by the first $\scriptA_k^l(t_{se}).N_a$ defenders as per the assignment $\beta_{c_k}^-$ and the rest defenders form the right group $\scriptD_k^r(t_{se})$ (see Fig.~\ref{fig:groupsAfterSplit}). We assign the defenders in $\scriptD_k^l(t_{se})$ only to the swarms in $\scriptA_k^l(t_{se})$ and those in $\scriptD_k^r(t_{se})$ only to the swarms in $\scriptA_k^r(t_{se})$. By doing so we may or may not obtain an assignment that minimizes the cost in \eqref{eq:MIQCQP_cost} but we reduce the computation time significantly and obtain a reasonably good assignment quickly. As in the function \assignHierarchical, the process of splitting is done recursively until the number of attackers' swarms is smaller than a pre-specified number $\underline{N}_{ac}(>2)$. The function \assignMIQCQP finds the defender-to-attacker-swarm assignment $\beta_{c}(t_{se})$ by solving \eqref{eq:defenders_to_attackers_and_swarm_assign_MIQCQP2} after setting $A_{uc}^{(k)}(t_{se})$ and $D_{c_k}^t(t_{se}^-)$ as empty sets, i.e., no assignments of the terminal defenders to the unclustered attackers as this assignment is already performed in the prior steps. 
    
    We have the following result about the worst-case computational cost of the hierarchical heuristic. 
    \begin{lemma} \label{lem:computation_cost_hierarchical}
        For a given assignment problem of assigning $|D_{c_k}(t_{se}^-)|$ defenders to $\scriptA_k(t_{se}).N_a \;( =|D_{c_k}(t_{se}^-)|)$ attackers divided into $\scriptA_k(t_{se}).N_{ac}$ clusters and $\scriptA_k(t_{se}).N_{uc}$ unclustered attackers with a given threshold $\underline{N}_{ac}(>2)$, the worst-case computational cost of the hierarchical heuristic in Algorithm \ref{alg:defenders_to_attackers_and_swarm_assign_Hierarchical} is:
        \be 
        \baa
        C_{H}^{comp}(t_{se},k)&\hspace{-2mm}=O(2^{(\scriptA_k(t_{se}).N_{uc})^2}+ (N_{rsM}' -1)2^{3\underline{N}_{ac}^2} \\
        &\quad + 2^{\underline{N}_{ac}n_{max}} + 2^{3n_{ac,k}^2})
        \eaa
        \ee 
        where $N_{rsM}' = \floor{\frac{\scriptA_k(t_{se}).N_{ac}}{\underline{N}_{ac}}}$, $n_{max}=(\scriptA_k(t_{se}).N_a - \scriptA_k(t_{se}).N_{uc} -3n_{ac,k}-3\underline{N}_{ac}(N_{rsM}'-1))$ and $n_{ac,k} = \scriptA_k(t_{se}).N_{ac} - N_{rsM}' \underline{N}_{ac}$.
    \end{lemma}
    \begin{proof}
    In Algorithm \ref{alg:defenders_to_attackers_and_swarm_assign_Hierarchical}, two CADAA problems (mixed integer quadratic programs) are solved (line 3 and 4) to assign the defenders to the left and right group of unclustered attackers. Suppose the number of unclustered attackers in left and right group are $N_{uc}^l = |\calA_{uc}^{(k),l}(t_{se})|$ and $N_{uc}^r = |\calA_{uc}^{(k),r}(t_{se})|$, respectively.

    Additionally, there are several rs-MIQCQPs that are solved in Algorithm \ref{alg:defenders_to_attackers_and_swarm_assign_Hierarchical} to assign defenders to the clusters of the attackers. Maximum number of the clusters in any rs-MIQCQP solved in Algorithm \ref{alg:defenders_to_attackers_and_swarm_assign_Hierarchical} is $\underline{N}_{ac}$. Based on the hierarchical breakdown of the original assignment problem, the maximum number of such rs-MIQCQP's is $N_{rsM}' = \floor{\frac{\scriptA_k(t_{se}).N_{ac}}{\underline{N}_{ac}}}$. Let $n_i \;(\ge 3 \underline{N}_{ac})$ denote the number of attackers in the $\underline{N}_{ac}$ clusters in the $i^{th}$ rs-MIQCQP for all $i \in \{1,2,3,...,N_{rsM}'\}$. Similarly, let $n_0$ be the number of attackers in the remaining $n_{ac,k} = \scriptA_k(t_{se}).N_{ac} - N_{rsM}' \underline{N}_{ac}$ clusters considered in a separate rs-MIQCQP. We also have that equal number of defenders are to be assigned to these attackers by solving these integer programs. Then, the worst-case computational cost of solving all integer programs in Algorithm \ref{alg:defenders_to_attackers_and_swarm_assign_Hierarchical} is:
    \be
    C^{comp} = O \bigl( \underbrace{2^{(N_{uc}^l)^2} + 2^{(N_{uc}^r)^2}}_{C_{uc}^{comp}} + \underbrace{2^{ n_0 n_{ac,k}} + \textstyle \sum_{i=1}^{N_{rsM}'} 2^{ n_i\underline{N}_{ac}}}_{C_c^{comp}} \bigr )
    \ee
    where $N_{uc}^l + N_{uc}^r=\scriptA_k(t_{se}).N_{uc}$, and $\sum_{i=0}^{N_{rsM}'} n_i = \scriptA_k(t_{se}).N_a - \scriptA_k(t_{se}).N_{uc}$. Since the assignments to unclustered and clustered attackers are made separately, we will find the maximum values of $C_{uc}^{comp}$ and $C_{c}^{comp}$ separately. 
    The maximum value of $C_{uc}^{comp}$ occurs when either $N_{uc}^l=\scriptA_k(t_{se}).N_{uc}$ and $N_{uc}^r = 0$ or $N_{uc}^l=0$ and $N_{uc}^r = \scriptA_k(t_{se}).N_{uc}$. 
    We have that $n_{ac,k}\le \underline{N}_{ac}$. Then, the maximum value of $C^{comp}$ subject to $\sum_{i=1}^{N_{rsM}'} n_i = \scriptA_k(t_{se}).N_a - \scriptA_k(t_{se}).N_{uc}$ occurs when all $n_i$, except one $n_i$ for some $i \in \{1,2,3,...,N_{rsM}'\}$, take their smallest values, i.e., when $n_0 = 3 n_{ac,k}$, $n_i = 3\underline{N}_{ac}$ for all $i \in \{2,3,..., N_{rsM}'\}$ and $n_1 = n_{max} = \scriptA_k(t_{se}).N_a - \scriptA_k(t_{se}).N_{uc} -  3 n_{ac,k} - 3\underline{N}_{ac} (N_{rsM}'-1)$. Hence, the worst-case computational cost of the hierarchical heuristic is $C_{H}^{comp}(t_{se},k) = O(2^{(\scriptA_k(t_{se}).N_{uc})^2}+ (N_{rsM}' -1)2^{3\underline{N}_{ac}^2} + 2^{\underline{N}_{ac}(\scriptA_k(t_{se}).N_a - \scriptA_k(t_{se}).N_{uc} -3n_{ac,k}-3\underline{N}_{ac}(N_{rsM}'-1))} + 2^{3n_{ac,k}^2})$. 
    \end{proof}
    
\subsection{Assignment when attackers' swarm does not avoid defenders}
When the attackers in a given swarm $\calA_{c_k}(t)$ do not try to avoid the defenders and instead just aim to reach the protected area, i.e., the attackers are risk-taking, then herding will not be an effective way of defense. Mathematically, this intention of swarm of attackers $\calA_{c_k}(t)$ to not avoid defenders and simply target protected area, is characterized by the following condition.
\be \label{eq:risk_taking_swarm}
\norm{\mathbf{r}_{ac_k} - \mathbf{r}_p} \le \norm{\mathbf{r}_{df_k}(0) - \mathbf{r}_p}  \; \& \; (\mathbf{r}_{ac_k} - \mathbf{r}_p)^T\mathbf{v}_{ac_k}  <0
\ee
This condition implies that the center of mass of attackers in $\calA_{c_k}(t)$ has come closer towards the protected area than the gathering center of the corresponding herding defenders in $\calD_{c_k}(t)$ and the attackers' average velocity vector points towards the protected area. In other words, the attackers in $\calA_{c_k}(t)$ are not necessarily moving away from the defenders and they intend to simply reach the protected area $\calP$, i.e., the attackers are risk taking.   
Once swarm $\calA_{c_k}(t)$ satisfies \eqref{eq:risk_taking_swarm}, the corresponding defenders $\calD_{c_k}(t)$ choose to intercept all the attackers in $\calA_{c_k}(t)$. The defenders in $\calD_{c_k}$ are assigned to intercept the attackers in $\calA_{c_k}$ by using CADAA similar to \eqref{eq:defender_attackers_assign_MIQP} with $A_{c_k}(t)$ and $D_{c_k}(t)$ at the place of $A_{uc}(0)$ and $I_d$, respectively.   

\subsection{Comparison of the assignment algorithms} \label{sec:computation_time_comparison}
In this section, we compare the computational performance of the assignment algorithms. Using the results from Lemma \ref{lem:computation_cost_MIQCQP2} and \ref{lem:computation_cost_hierarchical}, we have the following result about the computational cost of the MIQCQP, the rs-MIQCQP and the heuristic in Algorithm \ref{alg:defenders_to_attackers_and_swarm_assign_Hierarchical}.

\begin{theorem}
Let Assumption \ref{assum:attackers_split} hold and $1< \underline{N}_{ac}<N_{ac}$, then the worst-case computational costs $C_{M}^{comp}$, $C_{rsM}^{comp}$ and $C_{H}^{comp}$ of the MIQCQP, the rs-MIQCQP and the heuristic, respectively, satisfy:
$ C_{H}^{comp}(t_{se},k) < C_{rsM}^{comp}(t_{se},k) \le C_{M}^{comp}(t_{se},k) $.
\end{theorem}
\begin{proof}
From Lemma \ref{lem:computation_cost_hierarchical}, we have:
\be \label{eq:worst_time_complexity_compare_1}
\baa
C_{H}^{comp}\hspace{-1mm}&\hspace{-2mm}=C_{H}^{comp}(t_{se},k)\\
\hspace{-2mm}&\hspace{-2mm}=O\bigl(2^{(\scriptA_k(t_{se}).N_{uc})^2}+ (N_{rsM}' -1)2^{3\underline{N}_{ac}^2} \\
        &\quad + 2^{\underline{N}_{ac}n_{max}} + 2^{3n_{ac,k}^2}\bigr)\\
& \hspace{-2mm} \le O\bigl(2^{(\scriptA_k(t_{se}).N_{uc})^2 + 3\underline{N}_{ac}^2(N_{rsM}' -1) + \underline{N}_{ac}n_{max} + 3n_{ac,k}^2}\bigr)\\
&\hspace{3.3cm} (\because 2^{\imath} +2^{\jmath} \le 2^{\imath+\jmath}, \forall \imath, \jmath \ge 1 )\\
& \hspace{-2mm} \le O\bigl(2^{\bigl((\scriptA_k(t_{se}).N_{uc})^2  +  \underline{N}_{ac}(\scriptA_k(t_{se}).N_a - \scriptA_k(t_{se}).N_{uc}) \bigr)} \times \\
& \qquad 2^{-3\underline{N}_{ac}n_{ac,k} + 3n_{ac,k}^2} \bigr) \\
& \hspace{-2mm} \le O\bigl(2^{\bigl((\scriptA_k(t_{se}).N_{uc})^2  +  \underline{N}_{ac}(\scriptA_k(t_{se}).N_a  \bigr)} \bigr) \\
&\hspace{5cm} (\because n_{ac,k}\le \underline{N}_{ac} ) \\
&\hspace{-2mm} < O\bigl (2^{ \min(2\scriptA_k(t_{se}).N_{uc}, |\calD_{c_k}(t_{se}^-)|) \scriptA_k(t_{se}).N_{uc}} \times  \\
& \qquad  2^{|\calD_{c_k}(t_{se}^-)|\scriptA_k(t_{se}).N_{ac}} \bigr )\\
&\hspace{4.5cm} (\because 1<\underline{N}_{ac}< N_{ac} ) \\
& \hspace{-2mm} = C_{rsM}^{comp}(t_{se},k)
\eaa
\ee
Using \eqref{eq:worst_time_complexity_compare_1} and the result from Lemma \ref{lem:computation_cost_MIQCQP2}, we can establish: $ C_{H}^{comp}(t_{se},k) < C_{rsM}^{comp}(t_{se},k) \le C_{M}^{comp}(t_{se},k) $.

\end{proof}

{Next, we analyze the average computational performance of the assignment algorithms by numerically evaluating random assignment scenarios on a computer with 16 core Intel-i7 processor and 64 GB RAM using MATLAB. The computation time for random initializations of the players for different numbers of clusters of the attackers and different numbers of the unclustered attackers is shown in Figure~\ref{fig:comp_cost_MIQCQP}(a), and that for different numbers of attackers is shown in Figure~\ref{fig:comp_cost_MIQCQP}(b). Each data point in Fig.~\ref{fig:comp_cost_MIQCQP} is obtained by taking average of the computational costs for 30 random sets of initial conditions of the players for each of the possible configurations of the clusters for the given number of clusters and the total number of agents. As one can observe, the computation time for MIQCQP increases with increase in total number of attackers as well as number of unclustered attackers. Furthermore, even for $N_a=30$ and $N_{uc} = 8$, the MIQCQP in \ref{eq:defenders_to_attackers_and_swarm_assign_MIQCQP2} takes around 25 s, which is not real-time implementable. Similarly, we show the computation times for the rs-MIQCQP and the hierarchical heuristic in Figure~\ref{fig:comp_cost_rsMIQCQP} and \ref{fig:comp_cost_Hierarchical}, respectively. As one can observe, the computational time for the respective scenarios for the rs-MIQCQP is significantly smaller than that for the MIQCQP, but rs-MIQCQP could still be too slow for a real-time operation. The heuristic has even smaller computation time than the rs-MIQCQP and thus more suitable for real-time operation, see the Figure~\ref{fig:comp_cost_comparison} for better comparison.} 


We also compare the resulting cost of the heuristic, $cost_{H}$, against the optimal cost, $cost_{rsM}$, obtained by solving the rs-MIQCQP by calculating the percentage error $\% E = \frac{100|cost_{rsM}-cost_{H}|}{cost_{rsM}}$. As one can observe in Fig.\ref{fig:assignCostError_comparison} the percentage error $\%E$ is below $4\%$ for all the evaluated cases. This means that the proposed heuristic provides an assignment solution that is very close to the one obtained by rs-MIQCQP within a fraction of the time taken by rs-MIQCQP. {
\textcolor{blue}{The heuristic algorithm can be run at around 2-5 Hz for problems with up to 60 attackers and up to 24 individual risk taking attackers.} The analysis providing theoretical guarantees on the cost of the heuristic is left open for future research.

\begin{figure}
	\centering
	\includegraphics[width=1\linewidth,trim={.3cm .5cm 0cm .5cm},clip]{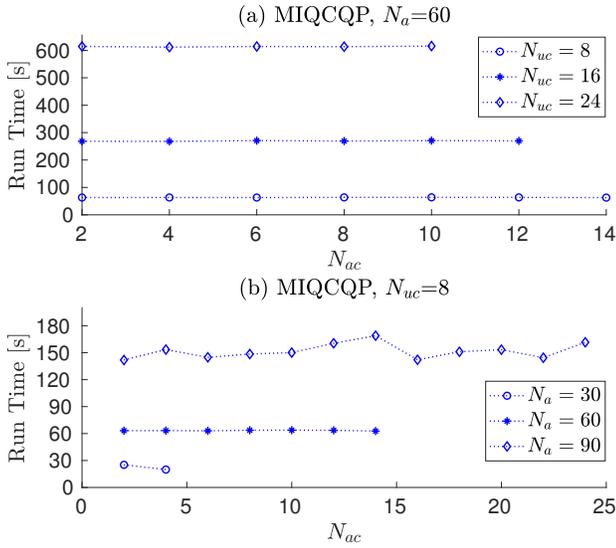}
	\caption{Computation time for MIQCQP in \eqref{eq:defenders_to_attackers_and_swarm_assign_MIQCQP}}
	\label{fig:comp_cost_MIQCQP}
\end{figure}

\begin{figure}
	\centering
	\includegraphics[width=1\linewidth,trim={.3cm .5cm 0cm .42cm},clip]{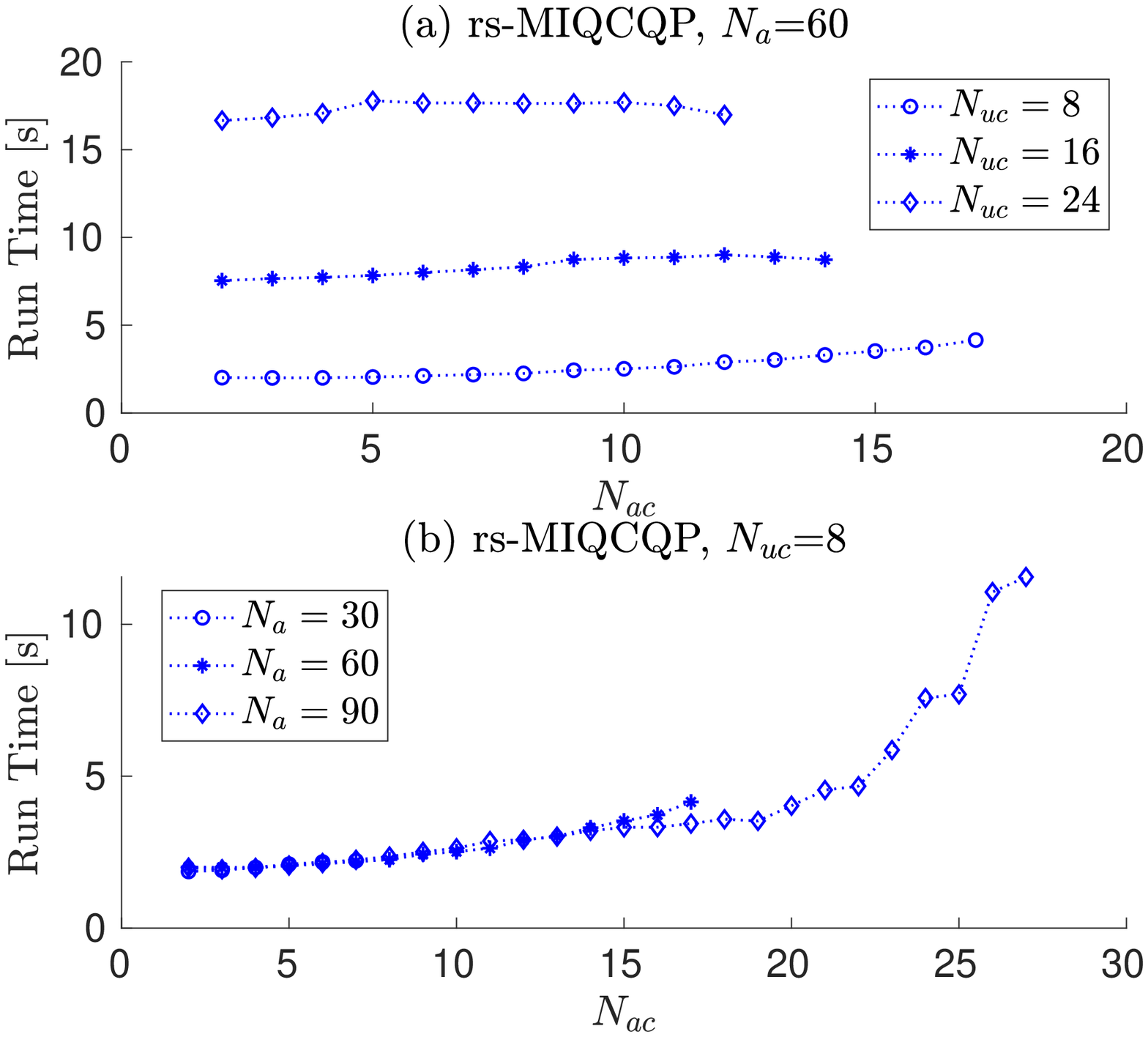}
	\caption{Computation time for rs-MIQCQP in \eqref{eq:defenders_to_attackers_and_swarm_assign_MIQCQP2}}
	\label{fig:comp_cost_rsMIQCQP}
\end{figure}

\begin{figure}
	\centering
	\includegraphics[width=1\linewidth,trim={.25cm .5cm 0cm .5cm},clip]{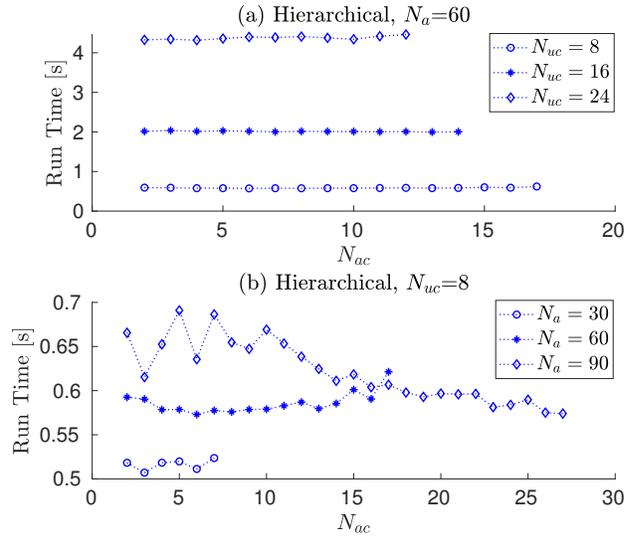}
	\caption{Computation time for Hierarchical Approach in Algorithm \ref{alg:defenders_to_attackers_and_swarm_assign_Hierarchical}}
	\label{fig:comp_cost_Hierarchical}
\end{figure}

\begin{figure}
	\centering
	\includegraphics[width=1\linewidth,trim={1.6cm .3cm 0cm .6cm},clip]{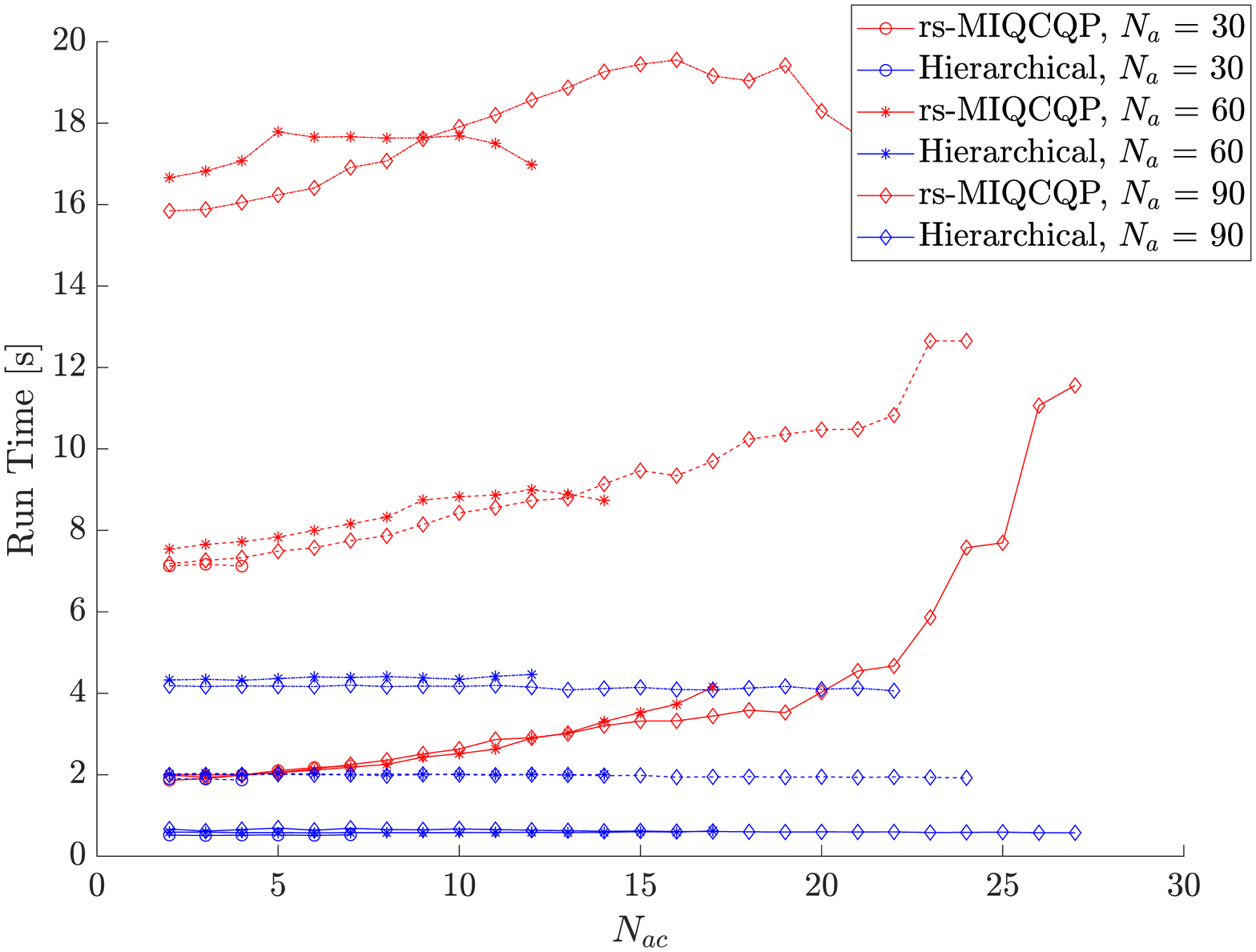}
	\caption{{Comparison of computation times of the rs-MIQCQP and the hierarchical heuristic (The line types solid (-), dash (-\hspace{-1mm} -), and dash-dot (-.) correspond to the cases with $N_{uc}=8$, $N_{uc}=16$, $N_{uc}=24$ respectively. 
			)}}
	\label{fig:comp_cost_comparison}
\end{figure}

\begin{figure}[h]
		\centering
		\begin{subfigure}[h]{1\linewidth}
		\includegraphics[width=1\linewidth,trim={.9cm 1.4cm 1.5cm 0cm},clip]{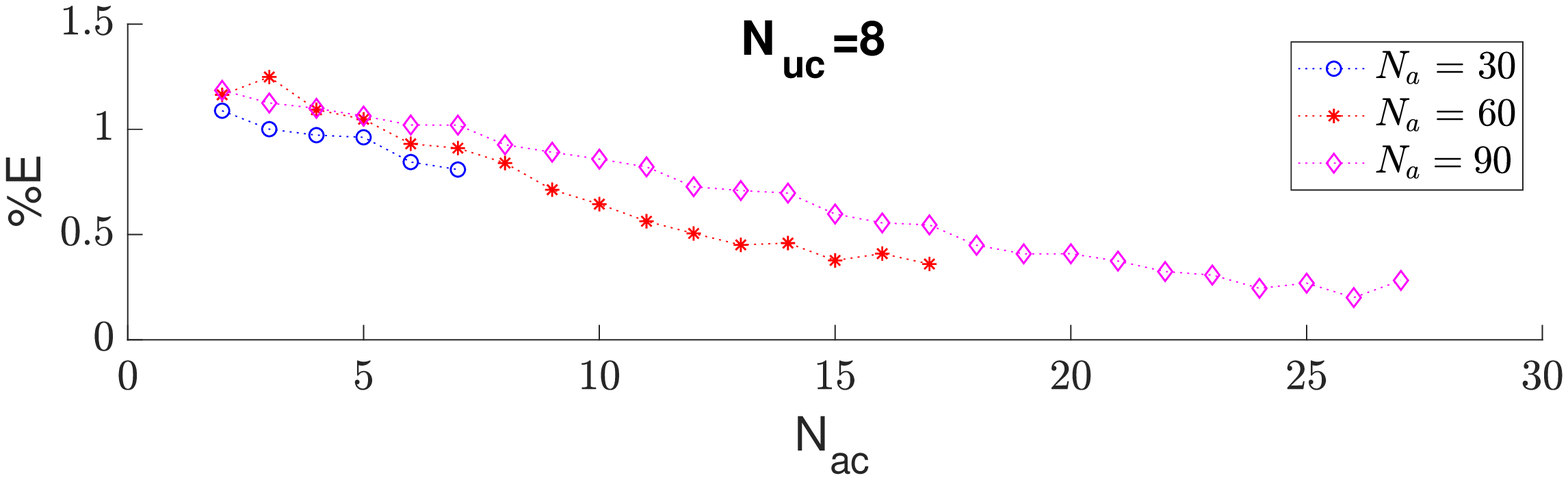}
			\label{fig:assignCostError_comparison1}
		\end{subfigure}	
		\begin{subfigure}[h]{1\linewidth}
		\includegraphics[width=1\linewidth,trim={.9cm 1.3cm 1.5cm 0cm},clip]{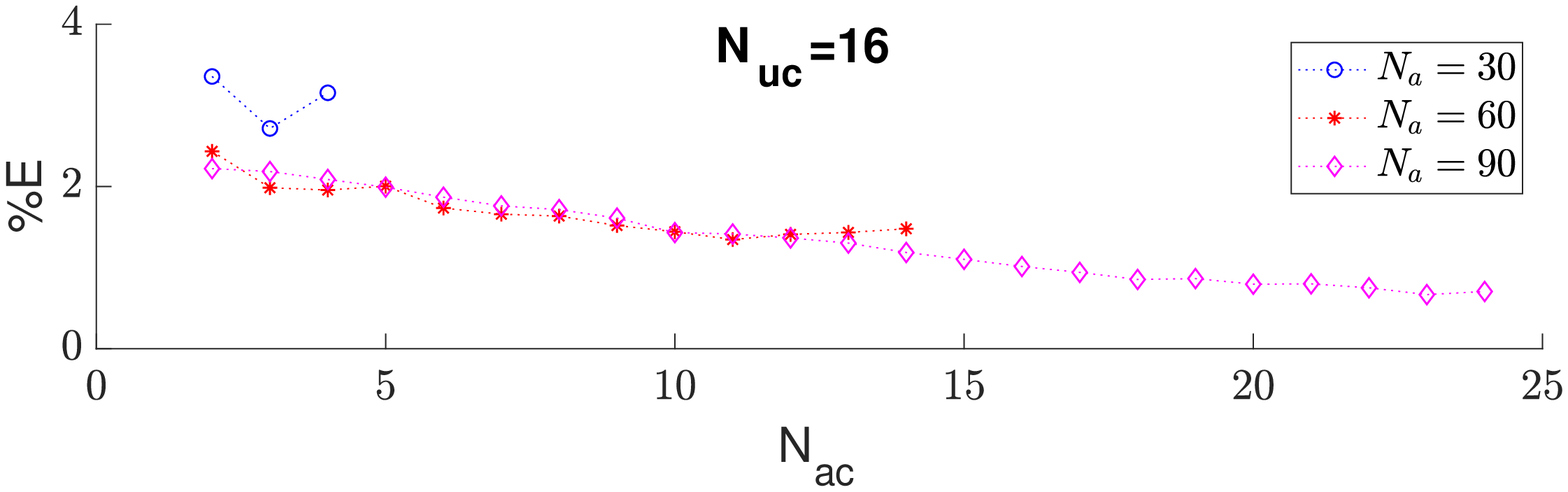}
		\label{fig:assignCostError_comparison2}
		\end{subfigure}
		\begin{subfigure}[h]{1\linewidth}
		\includegraphics[width=1\linewidth,trim={.9cm 0cm 1.5cm 0cm},clip]{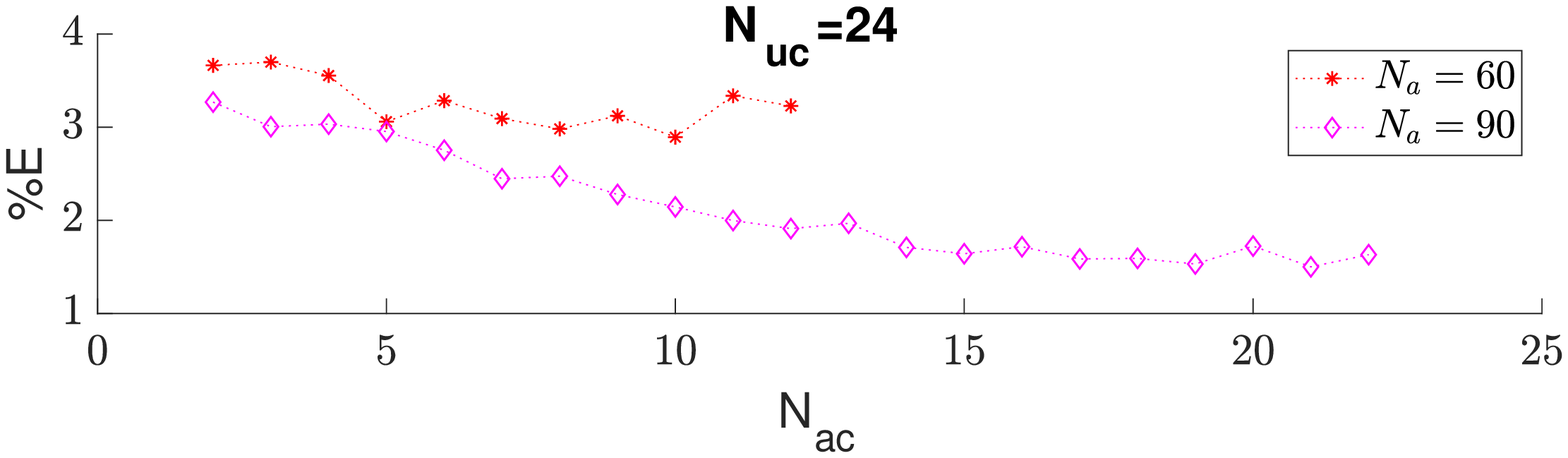}
		\label{fig:assignCostError_comparison3}
		\end{subfigure}
		\vspace{-5mm}
		\caption{\% Error in the costs of the rs-MIQCQP and the hierarchical heuristic}
		\label{fig:assignCostError_comparison}
\end{figure}


\subsection{Control augmentation for inter-defender collision avoidance}
The intercepting defenders need to avoid collisions with other intercepting as well as the herding defenders for their own safety. Each intercepting defender $\calD_j$, for all $j \in D_{uc}(t)$, employs an exponential CBF (ECBF) \cite{nguyen2016exponential,wang2017safe} based control augmentation to avoid collisions with other defenders such that their time-optimal control action corresponding to their assigned attacker is minimally augmented. This ECBF based control considers the Open-StringNets and Close-StringNets formed by the sub-teams of the herding defenders as big individual agents with their corresponding formation radii that the individual intercepting defenders need to avoid.

\section{Simulation Results}\label{sec:simulations}

In this section, we provide MATLAB simulations to demonstrate the effectiveness of the {multi-mode} defense strategy in different scenarios, as explained below. Some key parameters used in the simulations are: $\rho_a=\rho_d=0.5 \hspace{.5mm} m$, $C_D=1.5$, $\bar{v}_a=6 \hspace{.5mm} m/s \;(\bar{u}_a=9 \hspace{.5mm}  m/s^2)$, $\bar{v}_d=12.27 \hspace{.5mm}  m/s \; (\bar{u}_d = 18.4 \hspace{.5mm} m/s^2)$, $\varrho_{d}^{int}= 5 \hspace{.5mm} m $, $\rho_p=45 \hspace{.5mm} m$. {The computer specifications used to run these simulations are the same as those used in Section \ref{sec:computation_time_comparison}.} 

\textcolor{blue}{We consider a total number of five scenarios (case studies) whose simulation videos are available at (\href{https://youtu.be/cofhjqudT9U}{https://youtu.be/cofhjqudT9U})}. For the interest of space, in this section we provide plots of the simulation of Scenario 3. The description of all scenarios, as well as the detailed results of Scenario 3, are given in the following subsections.

{\subsubsection{Defenders and Attackers are equal in number} We consider three  different scenarios. 
\begin{itemize}
    \item Scenario (1): There are 32 attackers that appear, at $t=0$, to be divided into swarms $\calA_{c_1}(0) = \{\calA_i | i \in\{1,2,...,20\}\}, \calA_{c_2}(0)=\{\calA_{i}|i \in \{21,22,...,29\}\}$ and unclustered attackers $\calA_{uc}(0) = \{\calA_{30},\calA_{31}, \calA_{32}\}$ that are trying to reach the protected area, and 32 defenders that are aiming to prevent the attackers from doing so. In this scenario, after some time, $\calA_{c_1}$ splits into 3 smaller swarms and some of the terminal attackers from $\calA_{c_2}$ separate into individual risk-taking attackers. 
    \item Scenario (2): There are 20 attackers that are divided into swarms $\calA_{c_1}(0) = \{\calA_i | i \in\{1,2,...,12\}\}, \calA_{c_2}(0)=\{\calA_{i}|i \in \{13,14,...,17\}\}$ and unclustered attackers $\calA_{uc}(0) = \{\calA_{18},\calA_{19}, \calA_{20}\}$. In this scenario, some of the attackers from $\calA_{c_1}$ separate as individual risk-taking attackers.  
\item Scenario (3): At t=0, when the attackers are first identified, they are observed to be distributed as: 2 swarms $A_{c_1}(0)=\{\calA_{i} | i \in \{1,2,3,...,10\}\}$, $A_{c_2}(0)=\{\calA_{i} | i \in \{11,12,13,14\}\}$, and unclustered attckers $\calA_{uc} (0)=\{\calA_{15},\calA_{16}\}$.
\end{itemize}
}

{In the interest of space, we only discuss Scenario 3 in more detail here.} For the purpose of demonstration, the motion of the unclustered attackers is simulated under the time-optimal control to reach the protected area. The problem of finding the defenders' assignment to the attackers and the gathering formations is solved using Algorithm~\ref{alg:gathering_formations_centralized}. This results into two sub-teams of defenders $D_{c_1}(0)=\{\calD_{12},\calD_{10}, \calD_{16}, \calD_{14}, \calD_{8}, \calD_{7}, \calD_{9}, \calD_{13}, \calD_{1}, \calD_{2}\}$ and $D_{c_2}(0)=\{ \calD_{15}, \calD_{11}, \calD_{6}, \calD_{3}\}$ being assigned to gather on the time-optimal paths of $\calA_{c_1}(0)$ and $\calA_{c_2}(0)$, respectively, and 2 individual defenders $\calD_{4}$ and $\calD_{5}$ being assigned to intercept the unclustered attackers $\calA_{15}$ and $\calA_{16}$, respectively. Figure~\ref{fig:multimodeDefense_1_snap_1} shows the paths traversed by the players until all defenders' sub-teams gather at their respective desired formations, between the time interval $[0, 77.66]$ sec. As observed, both sub-teams of the defenders are able to successfully gather on the desired formations before respective attackers' swarm could reach there. The paths for the defenders in $\calD_{c_1}(0)$ and the attackers in $\calA_{c_1}(0)$ during the time interval $[77.66, 130.14]$ sec are shown in Figure~\ref{fig:multimodeDefense_1_snap_3}. As one can observe, the attackers $\calA_{c_1}(0)$ split at $t=t_{se} = 93.12$ sec into two smaller swarms $\calA_{c_1}(t_{se}) = \{\calA_{2}, \calA_{3}, \calA_{4}, \calA_{5}\}$ and  $\calA_{c_3}(t_{se}) = \{\calA_{6}, \calA_{7}, \calA_{8}, \calA_{9}\} $, and two outermost attackers, classified as unclustered attackers $\calA_{uc}^{(1)} (t_{se}) = \{\calA_{1}, \calA_{10} \}$, separate from the rest of the attackers in an attempt to circumvent the oncoming defenders. After solving the rs-MIQCQP \eqref{eq:defenders_to_attackers_and_swarm_assign_MIQCQP2}, the defenders in $\calD_{c_1}(0)$ are also divided into two smaller sub-teams $\calD_{c_1}(t_{se}) = \{\calD_{10}, \calD_{16}, \calD_{14}, \calD_{8}\}$ and  $\calD_{c_3}(t_{se}) = \{\calD_{7}, \calD_{9}, \calD_{13}, \calD_{1}\}$ and two terminal defenders $\calD_{12}$ and $\calD_{2}$. The sub-teams $\calD_{c_1}(t_{se})$ and $\calD_{c_3}(t_{se})$ are assigned to herd $\calD_{c_1}(t_{se})$ and $\calD_{c_3}(t_{se})$, respectively. And, the terminal defenders $\calD_{12}$ and $\calD_{2}$ are tasked to intercept the unclustered attackers $\calA_{1}$ and $\calA_{10}$, respectively. By the time $t = 130.14$ sec the two unclustered attackers are already captured and the two swarms of attackers are also completely enclosed by Closed-StringNets $\calG_{sn}^{cl}(\calD_{c_1}(t_{se}))$ and $\calG_{sn}^{cl}(\calD_{c_3}(t_{se}))$. Similarly, as shown in Figure~\ref{fig:multimodeDefense_1_snap_2} the defenders in $\calD_{c_2}(0)$ also successfully enclose the attackers in $\calA_{c_2}(0)$ at $t=146.17$ sec. Finally, as observed in Figure~\ref{fig:multimodeDefense_1_snap_4} all the enclosed attackers' swarms are herded to the respective closest areas by the Closed-StringNets formed by the defenders' sub-teams.  \textcolor{blue}{As mentioned also above, simulations for the additional scenarios are provided in the simulation video available at \href{https://youtu.be/cofhjqudT9U}{https://youtu.be/cofhjqudT9U}.}

{\subsubsection{Attackers outnumber the defenders} We also studied the performance of the proposed algorithm in a few scenarios where attackers outnumber the defenders. Particularly, we consider the following two scenarios. 
\begin{itemize}
    \item Scenario (4): There are 16 attackers that are, at $t=0$, divided into 2 swarms $\calA_{c_1}(0) = \{\calA_i | i \in\{1,2,...,6\}\}, \calA_{c_2}(0)=\{\calA_{i}|i \in \{7,8,...,14\}\}$ and unclustered attackers $\calA_{uc}(0) = \{\calA_{15},\calA_{16}\}$ and there are only 14 defenders. In this scenario, since the defenders are short in number by 2 and there are 2 swarms of attackers, resource allocation assigns 5 defenders ($\calD_{c_1}(0) = \{\calD_5, \calD_7, \calD_8, \calD_{12}, \calD_{13}\}$) to $\calA_{c_1}$ which has 6 attackers in it and 7 defenders ($\calD_{c_2}(0) = \{\calD_2, \calD_1, \calD_9, \calD_{6}, \calD_{10}, \calD_{11},\calD_{14}\}$) to $\calA_{c_2}$ which has 8 attackers in it and the remaining two defenders to intercept the unclustered attackers. As time progresses, at around $t_{se}=93.58$ sec, $\calA_{c_2}(t_{se}^-)$ splits into two smaller swarms $\calA_{c_2}(t_{se}) = \{\calA_7, \calA_8,\calA_9, \calA_{10}\}$ and $\calA_{c_3}(t_{se}) = \{ \calA_{11}, \calA_{12},\calA_{13}, \calA_{14} \}$. Again, since $\calD_{c_2}(t_{se}^-)$ is short by 1 defender, only 3 defenders ($\calD_{c_2}(t_{se}) = \{\calD_2, \calD_1, \calD_9\}$) are assigned to $\calA_{c_2}(t_{se}$ and 4 defenders ( $\calD_{c_3}(t_{se}) = \{\calD_{6}, \calD_{10}, \calD_{11},\calD_{14}\}$) are assigned to $\calA_{c_3}(t_{se})$. \textcolor{blue}{The trajectories of the players for this scenario are shown in the simulation video (\href{https://youtu.be/cofhjqudT9U}{https://youtu.be/cofhjqudT9U})}. As one can observer in the video, the defenders are still able to enclose the attackers' swarms successfully and herd them to respective safe areas despite more number of attackers in the attacking swarms. This is because the attackers did not disperse and stayed in compact formations throughout, that the available defenders were capable of enclosing with the given constraints ($\bar{R}_{}$). However, this is a very specific behaviour by the attackers that results in outcomes in favor of the defenders. 
\item Scenario (5): There are 6 attackers, all of them individual attackers and only 4 defenders. The four attackers ($\calA_1,
\calA_2,\calA_3,\calA_4$) approach the protected area from one side and the other two ($\calA_5,\calA_6$) approach the protected area from the opposite side. Because of the initial states of the defenders, ($\calD_2, \calD_4, \calD_3, \calD_1)$ are assigned to attackers ($\calA_1,
\calA_2,\calA_3,\calA_4$) in that order. After the defender $\calD_3$ and $\calD_1$ capture their target attackers they get assigned to $\calA_6$ and $\calA_5$ respectively. \textcolor{blue}{Again, the trajectories of the players are shown in the simulation video (\href{https://youtu.be/cofhjqudT9U}{https://youtu.be/cofhjqudT9U})}. As one can observe in the video, despite the re-assignment, the attackers $\calA_5$ and $\calA_6$ are able to reach the protected area. This is because the attackers $A_1-A_4$ started moving away from the protected area as they saw the defenders coming towards them. By the time the $\calD_3$ and $\calD_1$ intercepted $\calA_3$ and $\calA_4$, the defenders had already moved very far from the protected area and hence were not able to come back in time and intercept the remaining two attackers.
\end{itemize}
} 

{
These two scenarios show that the success of the defenders when attackers outnumber the defenders is not necessarily govern by the difference in their number but rather by the initial state of the players and how the attackers behave.
}

    \begin{figure*}[h]
		\centering
		\begin{subfigure}[h]{0.5\textwidth}
		\includegraphics[width=1\linewidth,trim={.1cm 0.1cm .2cm .25cm},clip]{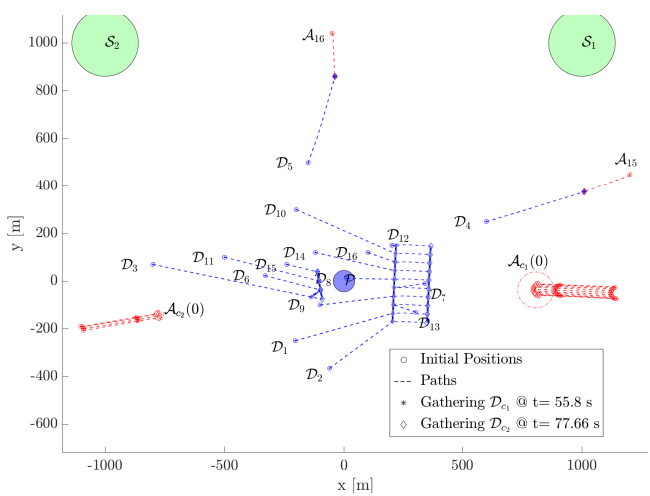}
			\caption{Paths of all agents during time interval $ [0, 77.66]$ s}
			\label{fig:multimodeDefense_1_snap_1}
		\end{subfigure}	
		\begin{subfigure}[h]{0.49\textwidth}
		\includegraphics[width=1\linewidth,trim={.5cm 0cm 1cm .25cm},clip]{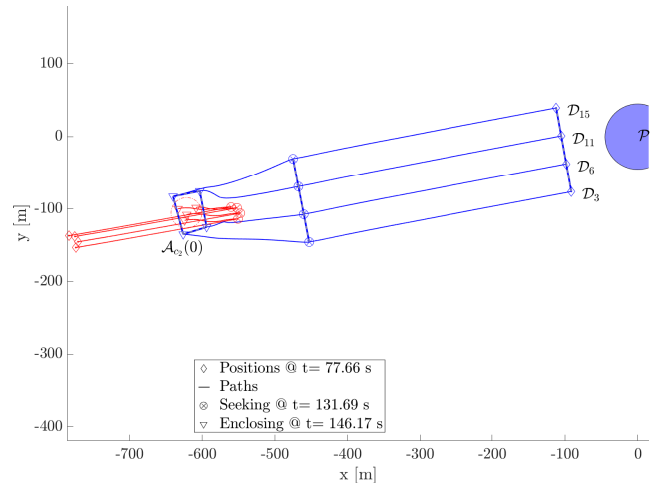}
		\caption{Paths of agents in $\calD_{c_2}(0)$ and $\calA_{c_2}(0)$ during time interval $ [77.66, 146.17]$ s}
		\label{fig:multimodeDefense_1_snap_2}
		\end{subfigure}
		\begin{subfigure}[h]{0.45\textwidth}
		\includegraphics[width=1\linewidth,trim={.1cm 0cm 6.5cm .25cm},clip]{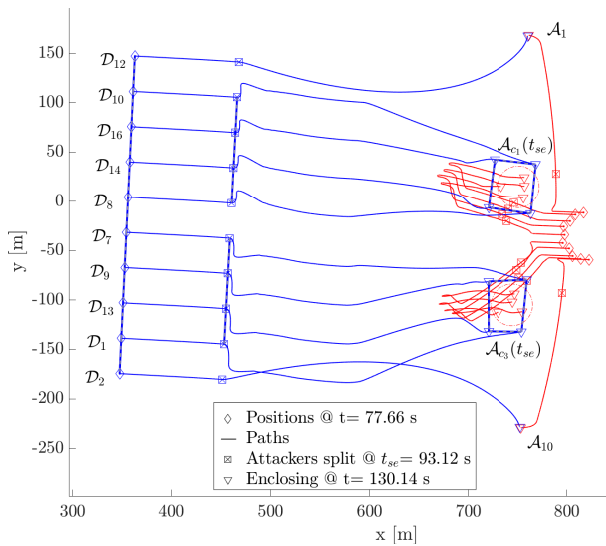}
		\caption{Paths of agents in $\calD_{c_1}(0)$ and $\calA_{c_1}(0)$ during time interval $ [77.66, 146.17]$ s}
		\label{fig:multimodeDefense_1_snap_3}
		\end{subfigure}
	\begin{subfigure}[h]{0.5\textwidth}
		\includegraphics[width=.9\linewidth,trim={.1cm 0cm 6cm .1cm},clip]{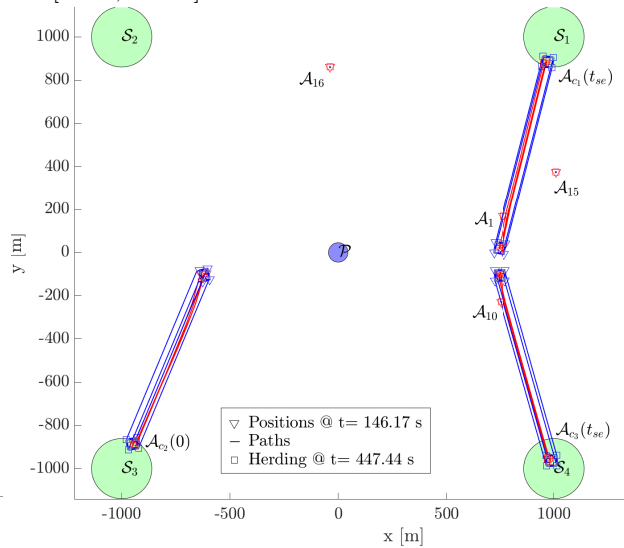}
		\caption{Paths of all agents during the herding phase (blue: defenders, red: attackers)}
		\label{fig:multimodeDefense_1_snap_4}
	\end{subfigure}
 		\caption{Snapshots of the paths of the agents during {multi-mode} defense (blue: defenders, red: attackers)}
		\label{fig:multimodeDefense1}
	\end{figure*}


\section{Conclusions} \label{sec:conclusions}
In this paper, we combine a {multi-mode} inter-defender collision-aware interception strategy (IDCAIS) with a swarm-herding strategy (StringNet Herding) to provide a {multi-mode} defense strategy against a wide range of behaviors by the attackers. We provided mixed-integer programs and computationally-efficient heuristics to allocate the interception or herding task to the defenders. Through simulations we showed how the defenders initially attempt to herd the attackers instead of intercepting the risk-averse swarms of the attackers, and how defenders redistribute to sub-teams and reassign either the herding or the interception role to themselves as the attackers split and take on risk-taking or risk-averse roles. The provided heuristics for solving the assignment problems offer a significant reduction in the computational time, {by at least a factor of 4-5}, while being close to the optimal solution, {within 4\% error}. Future work will focus on considering modeling and measurement uncertainty, as well as extending the formulation to 3D spaces.


	\vspace{-1mm}
	\bibliographystyle{IEEEtran}
	\bibliography{journal_interception_herding_Refs}
\end{document}